\documentclass[fleqn,usenatbib]{mnras}

\usepackage[T1]{fontenc}
\usepackage{ae,aecompl}
\usepackage[super]{nth}

\usepackage{graphicx} 
\usepackage{amsmath} 
\usepackage{amssymb} 
\usepackage{booktabs}
\usepackage{enumitem}
\usepackage[dvipsnames,table,xcdraw]{xcolor}
\usepackage{newtxtext,newtxmath}

\newcommand{\MSun}{$\mathrm{M}_{\odot}\mathrm{\ }$}

\newcommand{\MEarth}{$\mathrm{M}_{\oplus}$} 

\newcommand\SI[2]{{#1\,\textrm{#2}}}
\def\apgt{\ {\raise-.5ex\hbox{$\buildrel>\over\sim$}}\ }
\def\aplt{\ {\raise-.5ex\hbox{$\buildrel<\over\sim$}}\ }

\title[Circumstellar discs in young star-forming regions]{Evolution of circumstellar discs in young star-forming regions}

\author[Concha-Ram\'irez et al.]{
Francisca Concha-Ram\'irez,
Maite J. C. Wilhelm,
Simon Portegies Zwart\thanks{E-mail: spz@strw.leidenuniv.nl}
\\
Leiden Observatory, Leiden University, PO Box 9513, 2300 RA Leiden, The Netherlands\\
}

\date{Accepted XXX. Received YYY; in original form ZZZ}

\pubyear{2022}

\begin{document}
\label{firstpage}
\pagerange{\pageref{firstpage}--\pageref{lastpage}}
\maketitle

\begin{abstract}
The evolution of circumstellar discs is influenced by their
surroundings. The relevant processes include external
photoevaporation due to nearby stars, and dynamical truncations. The
impact of these processes on disc populations depends on the
star-formation history and on the dynamical evolution of the region.
Since star-formation history and the phase-space characteristics of
the stars are important for the evolution of the discs, we start
simulating the evolution of the star cluster with the results of
molecular cloud collapse simulations. 
In the simulation we form stars with circumstellar discs, which can
be affected by different processes. 
Our models account for the viscous evolution of the discs, internal and external 
photoevaporation of gas, external photoevaporation of dust, and dynamical
truncations. All these processes are resolved together with the
dynamical evolution of the cluster, and the evolution of the stars.

An extended period of star formation, lasting for at least 2\,Myr,
results in some discs being formed late.  These late formed discs have
a better chance of survival because the cluster gradually expands with
time, and a lower local stellar density reduces the effects of
photoevaporation and dynamical truncation.  Late formed discs can then
be present in regions of high UV radiation, solving the proplyd
lifetime problem. We also find a considerable fraction of discs that
lose their gas content, but remain sufficiently rich in solids to be
able to form a rocky planetary system.

\end{abstract}

\begin{keywords}
methods: numerical; planets and satellites: formation; stars: kinematics and dynamics;
\end{keywords}



\section{Introduction}

Circumstellar discs are a natural consequence of the star formation process and emerge within the first $10^4$ yr after star formation \citep{williams2011}. The star formation environment, rich in gas and newly-formed stars, can affect the evolution of the discs. The imprint of this environment on the young discs affects their potential to form planets and the topology of the eventual planetary system.

There are several ways in which the environment can influence the
evolution of circumstellar discs. Close encounters between
circumstellar discs and stellar fly-bys can affect the discs' size,
mass, and surface density. Close encounters can remove mass from the
outskirts of the discs, decreasing both their mass and radius
\citep[e.g.][]{clarke1991, clarke1993, pfalzner2005, breslau2014,
  vincke2015, vincke2016, portegieszwart2016, vincke2018, cuello2018,
  winter2018, concha-ramirez2019}. Several numerical implementations
of this process have shown that close encounters can lead to a less
steep disc's surface density \citep{rosotti2014}, the formation of
spiral arms and other structures \citep{pfalzner2003, pfalzner2005a},
accretion bursts onto the host star \citep{pfalzner2008}, and exchange
of mass between discs \citep{pfalzner2005, jilkova2016}. Observational
evidence for the effects of stellar fly-bys has been presented in
several studies. \citet{cabrit2006} study the $\sim 600$\,au trailing
"tail" in the disc of RW Aur A and argue that a recent fly-by might
have caused it. \citet{reche2009} demonstrate that the spiral arms
observed in the disc of the triple star system HD 141569 might be the
result of a fly-by.

The observed structure in GW Ori's circum-triple disc \citep{2021MNRAS.508..392S} may have been caused either by resonances in dynamics or by planet formation.
Observations by \citet{rodriguez2018} reveal newly-detected tidal streams in RW Aur A, and they propose that these might be the result of many subsequent close encounters. \citet{winter2018b} simulate the disc around DO Tau, which presents a tidal tail, and argue that this shape could have been caused by a close encounter with the nearby triple system HV Tau. There is also evidence that the young disc of the solar system was affected by such an encounter. The sharp edge of the solar system at $\sim 50$\,au could be a sign that a passing star truncated its early disc \citep{breslau2014, punzo2014}. The highly eccentric and inclined orbits of the \emph{Sednitos}, a group of 13 detected planetoids in the outskirts of the solar system, suggest they might have been captured from the disc of a nearby passing star \citep{jilkova2015}.

Another mechanism that can alter the evolution of circumstellar discs
is photoevaporation, through which high-energy photons heat the discs'
surfaces, causing it to evaporate. The source of these photons can
be the host star (internal photoevaporation) or bright stars in the
vicinity (external photoevaporation). Photoevaporation is driven by
far-ultraviolet (FUV), extreme ultraviolet (EUV), and X-ray photons
\citep{johnstone1998, adams2004}. The effects of internal and external
photoevaporation on circumstellar discs are distinct. Internal
photoevaporation can clear areas of the disc at specific disc radii,
causing the opening of gaps \citep{gorti2009, gorti2009a, owen2010,
  font2004, fatuzzo2008, hollenbach2000}. External photoevaporation
can remove mass from all over the disc surface, but the outer regions
of the discs are more vulnerable because the material is less strongly
bound to the host star \citep{johnstone1998, adams2004, haworth2019}.

Observational evidence of external photoevaporation was first obtained through the imaging of evaporating discs in the Orion nebula \citep{odell1994, odell1998}. These objects, now known as `proplyds', are circumstellar discs immersed in the radiation fields of nearby stars. Their cometary tail-like structure reveals the ongoing mass loss. Subsequent observations of the region showed that circumstellar disc masses decrease when close to massive stars. This effect has been observed in several regions such as the Trapezium cluster \citep[e.g.][]{vicente2005, eisner2006, mann2014}, the Orion Nebula Cluster \citep[e.g.][]{mann2010, eisner2018}, Cygnus OB2 \citep{guarcello2016}, NGC 1977 \citep{kim2016}, NGC 2244 \citep{balog2007}, Pismis 24 \citep{fang2012}, NGC 2024 \citep{vanterwisga2020}, $\sigma$ Orionis \citep{ansdell2017}, and $\lambda$ Orionis \citep{ansdell2020}. Younger and low-mass star-forming regions such as Lupus, Taurus, Ophiuchus, and the Orion Molecular Cloud 2 tend to have higher average disc masses than denser regions such as the Orion Nebula Cluster \citep{eisner2008, ansdell2016, eisner2018, vanterwisga2019}. \citet{vanterwisga2020} present the discovery of two distinct disc populations, in terms of mass, in the NGC 2024 region. The discs to the east of the region are embedded in a dense molecular ridge and are more massive than the discs outside the ridge, which are also closer to two OB-type stars. They propose that the difference in masses is caused by the eastern population being protected from the radiation of the nearby massive star IRS 1.

Several models have demonstrated that external photoevaporation is
efficient in depleting disc masses on timescales much shorter than
their estimated lifetimes of $\sim 10$\,Myr \citep[e.g.][]{scally2001,
  adams2006, fatuzzo2008, haworth2016}, even in low radiation fields
\citep{facchini2016, kim2016, haworth2017}. Because external
photoevaporation is caused by massive stars in the vicinity, the
extent of its effect depends on the density of the stellar region and
the number of massive stars in the surroundings. In high-density
regions ($\mathrm{N}_* \gtrsim 10^4$ pc$^{-3}$) where the stellar
population follows a well-sampled IMF, the disc mass-loss rates caused
by external photoevaporation are orders of magnitude higher than those
caused by dynamical truncations \citep{winter2018a, winter2019a,
  concha-ramirez2019a}. \citet{concha-ramirez2021} show that, in
regions of local stellar densities $\mathrm{N}_* > 100
\mathrm{\ pc}^{-3}$, external photoevaporation can evaporate up to
90\% of circumstellar discs within 2.0\,Myr. In low density regions
($\sim 10 \mathrm{\ M}_{\odot} \mathrm{\ pc}^{-3}$), only $\sim 60\%$
of discs are evaporated within the same timescale. \citet{winter2020a}
model a region comparable to the central molecular zone of the Milky
Way (surface density $\Sigma_0 = 10^3 \mathrm{\ M}_{\odot}
\mathrm{\ pc}^{-2}$) and find that external photoevaporation destroys
90\% of circumstellar discs within 1.0\,Myr. In regions of lower
density ($\Sigma_0 = 12 \mathrm{\ M}_{\odot} \mathrm{\ pc}^{-2}$) they
find a mean disc dispersal timescale of 3.0\,Myr. Similar results are
obtained by \citet{nicholson2019} who find that external
photoevaporation destroys 50\% of discs within \SI{1.0}{Myr} in
regions of density $\sim 100 \mathrm{\ M}_{\odot} \mathrm{\ pc}^{-3}$,
and within \SI{2.0}{Myr} in regions of density $\sim 10
\mathrm{\ M}_{\odot} \mathrm{\ pc}^{-3}$.

While observational and numerical evidence indicate that disc masses
decrease with increasing stellar density, massive discs are still
observed in high-density regions. For example in the ONC, which
contains discs with a mass higher than those in the proximity of the
massive star $\theta^1$ Ori C. If the discs are coeval with $\theta^1$
Ori C, they should have already been dispersed by external
photoevaporation, unless they were extraordinarily massive to begin
with ($\mathrm{M}_\mathrm{disc} \gtrsim 1
\mathrm{\ M}_{\odot}$). Alternatively, $\theta^1$ Ori C would have to
be considerably younger than the ONC average ($\lesssim
\SI{0.1}{Myr}$) for these discs to have survived. This is known as the
`proplyd lifetime problem'. \citet{storzer1999} propose that these
discs are currently passing through the region's center, whereas they
spent most time at a larger distance, where the radiation is
ineffective. \citet{scally2001} model external photoevaporation on a
cluster similar to the ONC and find that the necessary radial orbits
proposed by \citet{storzer1999} are not dynamically plausible in such
a region. \citet{winter2019a} revisit the problem and propose a
solution to describe why these discs exist. Their solution consists of
a combination of factors: different eras of star formation allow for
massive discs to be around stars that are younger than the average of
the ONC population; stars forming in subvirial states with respect to
the gas potential allows young stars to migrate to the central region
of the ONC; and interstellar gas protects the discs from the
radiation, allowing them to live longer than expected.

The star-formation history and primordial stellar distributions in
young star-forming regions are key to understanding the environment's
effects on the disc populations. The star formation process results in
regions with different morphologies, with clumps and filaments likely
to be present. This structure is far from the spherical, idealized
initial conditions commonly used in models of star clusters. The
collapse of the giant molecular clouds (GMCs) from which stars form is
affected by turbulent flows \citep{falgarone1991, falgarone1991a}
which result in filamentary, clumpy, or fractal gas substructure in
the cloud \citep[e.g.][]{scalo1990, larson1995, elmegreen2000,
  hacar2018}. The distribution of the newly-formed stars follows the
local densities of the gas in the molecular clouds. Star clusters
would then form in a bottom-up scenario, where subclusters emerge from
clumps and filaments with high star-formation efficiency that
eventually merge \citep[see]{2016ApJ...817....4F}. This has been
determined theoretically \citep[e.g.][]{elmegreen2008, kruijssen2012a}
and suggested by recent observations \citep[e.g.][]{ward2020}. Other
authors \citep[e.g.][]{banerjee2015, kuhn2019} have proposed a
monolithic or top-down formation scenario for star clusters, i.e., one
where a single, highly active star-formation episode yields a compact
cluster that remains embedded in its primordial gas. Once feedback
processes expel this gas, the cluster expands and loses stars. While
circumstellar discs emerge during the protostellar phase
\citep{williams2011}, the star formation process defines the
environment in which the discs are immersed at their early stages.

It is important to take a step back in time to better understand the
environmental effects on circumstellar discs and study how the star
formation process influences stellar densities.  This work presents a
model for circumstellar discs inside young star-forming regions. We
adopt a relatively simple model for the star formation process,
starting from the collapse of a giant molecular cloud to obtain
masses, positions, and velocities of newly formed stars. These form
the input for our star cluster evolution code. During the evolution,
we take into account the viscous evolution of the discs, dynamical
truncations, external and internal photoevaporation of gas, and external
photoevaporation of dust. We evolve the discs simultaneously with the stellar
dynamics and stellar evolution.

\section{Model}

We simulate the formation of massive clusters of bound stars, including circumstellar discs. These discs are subject to viscous spreading, dynamical truncations, and photoevaporation.

We model several different astrophysical processes which operate
simultaneously on a range of scales: the collapse of a molecular
cloud, star formation, stellar dynamics, and viscous circumstellar
discs, which are affected by dynamical truncations and
photoevaporation. We bring these processes together using the
Astrophysical Multipurpose Software Environment, AMUSE
\citep{portegieszwart2013, pelupessy2013}. The results presented in
this work are obtained through two different simulation stages: first,
we simulate the collapse of a molecular cloud, including star
formation. This results is a distribution of stars in space with ages,
masses and velocities. We continue to evolve these stars in the second
stage, including the evolution of the circumstellar discs. This second
stage encompasses the stellar dynamics, stellar evolution, viscous
evolution of the discs, and photoevaporation. All the code developed
for this work is available
online.

\subsection{Stage 1: The collapse of the molecular cloud and the formation of stars}\label{model:cloud}

In the first stage, we model the star-formation process. We simulate the collapse of a molecular cloud using the smoothed particle hydrodynamics (SPH) code \texttt{Fi} \citep{pelupessy2004}. The simulations start with a spherical gas reservoir of mass $10^4$ \MSun with an initial radius of \SI{3}{pc}. The SPH simulations use 32.000 equal-mass particles, with a mass resolution of 0.3 \MSun per particle. The softening in the simulations is $\epsilon = 0.05$\,pc. A power-law velocity spectrum $P(k) \propto k^{-4}$ was adopted to emulate large scale turbulence (here $k$ is the wavenumber \citep{bate2003}). These initial conditions result in a velocity dispersion in three dimensions that scales with the observed relations by \citet{larson1981}. The mean temperature of the gas is 23 K, and the initial free-fall timescale is \SI{0.42}{Myr}.

We model the star formation process using sink particles, which are
created from regions of the cloud where the local gas density is
higher than $1 \mathrm{M}_\odot / {\epsilon}^3 \simeq
8000$\,M$_\odot$/pc$^3$. Once formed, sink particles continue to
accrete gas from the molecular cloud. Each of these sink particles can
form several stars.

Since we aim to preserve a power-law stellar mass function similar to the one observed in the galaxy, the stars in our simulations are formed by introducing a predetermined star-formation efficiency (SFE). We set a SFE of 0.3 \citep{lada2003}. In reality, the integrated SFE of molecular clouds is much lower, and values around 0.3 can be found around much smaller cores \citep[e.g.][]{matzner2000, chevance2020a, smith2020}. 

We implement this SFE by keeping track of the total initial mass of all stars. When this mass exceeds 30\% of the initial cloud mass, the SPH code is stopped and all remaining gas and sink particles are instantaneously removed. 

The star formation process begins once sink particles have accreted
enough mass to sample stars randomly from the initial mass function
(IMF). We base the star formation mechanism on \citet{wall2019}. We
begin by sampling a random stellar mass $m$ from a Kroupa IMF
\citep{kroupa2001} of 10.000 stars, with a lower limit of 0.08 \MSun
and upper limit 150 \MSun \citep{wall2019}. Then, the list of sinks is
checked to find one that is massive enough to form the next star in
the list. The first sink to satisfy this condition will be selected.
We subtract the mass $m$ from the sink, and a new star is born. If $m
\leq 1.9 \mathrm{M}_{\odot}$ the star will have a circumstellar disc
(see section \ref{model:extphotoevap}), and we subtract the mass of
the star and the initial mass of the disc from the sink. The position
of the newly formed star is determined by taking the position of the
sink, and we add a random offset in each spatial dimension. This offset is
uniformly distributed within plus and minus the sink radius. The sink
radius corresponds to the distance from which the sink accretes
material. By the time star formation has started, sink radii are
$8\cdot 10^3$\,au. The velocity of the new star is set to the velocity
of its birth sink.

After a sink has formed a star, we set a delay time that must pass
before creating a new star. We implement this step to prevent the
instantaneous conversion of gas into stars. This is typical if only
hydrodynamics and self-gravity are accounted for, but observations
indicate that star formation proceeds at a lower rate
\citep{krumholz2019}. Various processes can contribute to this, such
as magnetic fields and stellar feedback, which we do not simulate.
The delay is implemented as an exponentially decaying timescale of

\begin{equation}\label{eq:delay}
\mathrm{t}_{\mathrm{delay}} = \mathrm{t}_{\mathrm{ff}} \mathrm{\ exp}\left(\frac{-\mathrm{\ t}}{t_d}\right).
\end{equation}

\noindent
Here $\mathrm{t}_{\mathrm{ff}}$ is the free-fall time scale of the corresponding sink, $t$ is the current model time, and $t_d$ is a free parameter we fix at \SI{1}{Myr}. {\bf It is not clear a priori how this recipe translates to a star formation rate, but we explore this in section \ref{sec:sfr}.}

During the molecular cloud-collapse simulation, we keep track of the
newly formed stars' mass, position, velocity, and birth time. This
data will then be input for the second part of the simulations, in
which the discs are evolved together with the stellar dynamics. The
dynamical evolution of the stars is calculated using the
$4^{\mathrm{th}}$-order Hermite integrator \texttt{ph4}, which is
integrated together with the SPH integrator in a leap-frog scheme
using the \texttt{Bridge} \citep{fujii2007} coupling method in AMUSE
\citep[see][for implementation details]{portegieszwart2020}.  Stellar
evolution was accounted for using the {\tt SeBa} stellar and binary
evolution package \citep{1996A&A...309..179P,2020A&A...640A..16T}.

\subsection{Stage 2: The dynamics of stars and the evolution of their circumstellar discs}\label{model:stage2}

The second simulation stage begins when the first star has formed. For
each star, we evolve its disc and calculate its external
photoevaporation mass-loss rate as explained in sections
\ref{model:discs} and \ref{model:extphotoevap}, respectively. The star
formation process ends when 30\% of the initial cloud mass has been
converted into stars. Then, the SPH code and \texttt{Bridge} coupling
between gas and stars are stopped, in effect instantaneously removing
the excess gas. From then on we only deal with the stars' dynamics and
evolution, and the processes as explained in the following
sections. This second stage of the simulations run for \SI{2}{Myr}
after the last star has formed. Below we describe how each of these
process is incorporated.

\subsubsection{Circumstellar discs}\label{model:discs}

We model circumstellar discs using the Viscous Accretion Disk [sic]
Evolution Resource \citep[VADER,][]{krumholz2015}. VADER models the
viscous transport of mass and angular momentum in thin, axisymmetric
discs. Each disc is defined with a grid of 100 logarithmically spaced
cells, spanning 0.05 to \SI{2000}{au}. The larger outer limit allows
the discs to expand without reaching the grid boundaries. The mass
flow through the outer boundary of the grid is set to zero to maintain
the density required to measure the disc radius. The mass flow from
the disc's inner boundary is considered accreted onto the host
star. Each disc has a Keplerian rotation profile and turbulence
parameter $\alpha = 5\cdot 10^{-3}$, which results in a viscous
timescale of $\sim \SI{0.1}{Myr}$.

We use the standard disc profile of \citet{lynden-bell1974} to establish the initial column density of the discs as:

\begin{equation}
\Sigma(r, t=0) = \frac{m_d}{2 \pi r_d \left(1 - e^{-1}\right)} \frac{\exp(-r/r_d)}{r}.
\end{equation}

\noindent
Here $r_d$ is the initial disc radius, $m_d$ is the initial disc
mass. We consider the characteristic radius to be $r_c \approx r_d$
\citep{anderson2013}.

For the external photoevaporation process, we keep track of the outer
edge of the disc.  We define the disc radius $r_d$ as the radius which
encloses 99\% of the disc mass \citep{anderson2013}, and set the
column density outside $r_d$ to a negligible value
$\Sigma_{\mathrm{edge}} = 10^{-12}\,g/cm^3$. The mass loss due to
external photoevaporation (section \ref{model:extphotoevap}), as well
as dynamical truncations (section \ref{model:truncations}), causes the
disc to develop a steep density profile at the outer edge. The
location of the edge is insensitive to the value of
$\Sigma_{\mathrm{edge}}$, given that it is sufficiently low
\citep{clarke2007}.

\subsubsection{External photoevaporation}\label{model:extphotoevap}

We calculate the mass loss due to external photoevaporation using the
Far-ultraviolet Radiation Induced Evaporation of Discs (FRIED) grid
\citep{haworth2018}. This grid provides a pre-calculated set of
mass-loss rates for discs immersed in UV radiation fields of varying
strengths, from $10 \mathrm{\ G}_0$ to $10^{4} \mathrm{\ G}_0$, where
$\mathrm{G}_0$ is the FUV field measured in Habing units, $1.6\cdot
10^{-3}$ erg/s/cm$^3$ \citep{habing1968}.  The grid spans discs of
mass $\sim10^{-4} \mathrm{\ M}_\mathrm{Jup}$ to $10^{2}
\mathrm{\ M}_\mathrm{Jup}$, radius from $\SI{1}{au}$ to
$\SI{400}{au}$, and host star mass from $0.05 \mathrm{\ M}_{\odot}$ to
$1.9 \mathrm{\ M}_{\odot}$.

To stay within the boundaries of the FRIED grid, we consider only stars of $\mathrm{M}_* \leq 1.9$ \MSun to have a circumstellar disc. More massive stars are considered to generate too much radiation to have an appreciable disc.
This mass distinction is for external photoevaporation calculations only; for the stellar dynamics evolution, there is no separation between these two stellar groups.

Far-ultraviolet (FUV) photons dominate in the external photoevaporation process \citep{armitage2000, adams2004, gorti2009a}. We calculate the FUV radiation from the massive stars by pre-computing a relation between stellar mass and FUV luminosity using the UVBLUE spectral library \citep{rodriguez-merino2005}. The obtained fit is presented in Figure 2 of \citet{concha-ramirez2019a}. We use that fit to determine the FUV radiation emitted by each massive star at every simulation time step.

The external photoevaporation process is applied at every time step. We first calculate the distance from every disc to every star of mass $\mathrm{M}_* > 1.9$ \MSun and determine the total radiation received by each disc. We do not consider extinction due to interstellar material. We interpolate from the FRIED grid using the calculated total radiation and the disc parameters to find a photoevaporation-driven mass loss rate for each disc. Assuming the mass loss rate $\dot{\mathrm{M}}$ to be constant during the current time step, we use it to calculate the total mass loss. This mass is subsequently removed from the disc's outer region, removing mass from each subsequent cell until a finite amount of mass remains. External photoevaporation then results in a decrease of disc mass and disc radius.

External photoevaporation can be dominated by extreme ultraviolet (EUV) photons. This happens when a disc is closer to a radiating star than \citep{johnstone1998}
\begin{equation}\label{eq:dmin}
d_{min} \simeq 5 \times 10^{17} \left(\frac{\varepsilon^2}{f_r
  \Phi_{49}}\right)^{-1/2} \textrm{r}_{d_{14}}^{1/2} \mathrm{\ cm}.
\end{equation}

\noindent
Here $f_r$ is the fraction of EUV photons absorbed in the ionizing
flow, $\Phi_{49} = ({\Phi_i}/{10^{49}}) \mathrm{s}^{-1}$ is the EUV
luminosity of the source, $\varepsilon$ is a dimensionless normalizing
parameter, $({\varepsilon^2}/({f_r \Phi_{49}}))^{1/2} \approx 4$, and
$r_{d_{14}} ={r_d}/({10^{14} \mathrm{cm}})$ with $r_d$ the disc radius.

When the distance $d$ between a disc and a massive star is $d < d_{min}$, the disc enters the EUV-dominated photoevaporation regime. The mass loss in this case is calculated as:

\begin{equation}\label{eq:euv}
\dot{M}_{EUV} = 2.0 \times 10^{-9} \frac{(1 + x)^2}{x} \varepsilon r_{d_{14}} \mathrm{\ M}_{\odot} \mathrm{\ yr}^{-1},
\end{equation}

\noindent
with $x \approx 1.5$ and $\varepsilon \approx 3$ \citep{johnstone1998}.

A disc is considered dispersed when it has lost 99\% of its initial
gas mass \citep{ansdell2016} or when its mean column density drops
below $1 g/cm^2$ \citep{pascucci2016}. After a disc is dispersed, its
host star continues to be integrated by the stellar dynamics and
stellar evolution codes.

\subsubsection{Internal photoevaporation}\label{model:intphotoevap}

Internal photoevaporation is driven by X-ray radiation \citep{owen2010, owen2012}. We calculate the X-ray luminosity of the stars with discs using the fit obtained by \citet{flaccomio2012} for T-Tauri stars as a function of stellar mass:

\begin{equation}
\log\left(\frac{L_X}{\mathrm{erg} \mathrm{\ s}^{-1}}\right) = 1.7 \log \left(\frac{\mathrm{M}_*}{1 \mathrm{\ M}_{\odot}}\right) + 30.
\end{equation}

\citet{picogna2019} calculate X-ray mass loss profiles and mass-loss rates for a star of mass 0.7 \MSun. \citet{owen2012} developed scaling relations that allow to calculate these values for stars of $\mathrm{M}_* \leq 1.5 \mathrm{M}_{\odot}$. We combine the results of \citet{picogna2019} with the scaling relations of \citet{owen2012} to span a larger range of stellar masses. The internal photoevaporation mass loss rate is then given by

\begin{equation}
\dot{M}_X = 10^{\Delta} \left(\frac{\mathrm{M}_*}{0.7 \mathrm{M}_{\odot}}\right)^{-0.068} \mathrm{\ M}_{\odot} \mathrm{\ yr}^{-1},
\end{equation}

\noindent
where

\begin{equation}
\Delta = -2.7326 \mathrm{\ exp}\left[\frac{(\ln (\log(L_X)) - 3.3307)^2}{2.9868 \times 10^{-3}}\right] - 7.2580
\end{equation}

\noindent
is the X-ray mass loss rate derived by \citet{picogna2019}.

The mass loss profile takes the form

\begin{multline}\label{eq:masslossprofile}
\dot{\Sigma}_w(x) = \ln(10) \Bigg(\Bigg.\frac{6 a \ln(x)^5}{x \ln(10)^b} + \frac{5 b \ln(x)^4}{x \ln(10)^5} + \frac{4 c \ln(x)^3}{x \ln(10)^4} + \frac{3 d \ln(x)^2}{x \ln(10)^3} \\+ \frac{2 e \ln(x)}{x \ln(10)^2} + \frac{f}{x \ln(10)} \Bigg.\Bigg) \frac{\dot{M}_w(x)}{2 \pi x} \mathrm{\ M}_{\odot} \mathrm{\ au}^{-2} \mathrm{\ yr}^{-1},
\end{multline}

\noindent
where

\begin{equation}
\dot{M}_w(x) = \dot{M}_X 10 ^{a\log x^6 + b\log x^5 + c\log x^4 + d\log x^3 + e\log x^2 + f\log x + g}.
\end{equation}

\noindent
Here $a = -0.5885$, $b = 4.3130$, $c = -12.1214$, $d = 16.3587$, $e = -11.4721$, $f = 5.7248$, and $g = -2.8562$ \citep{picogna2019}, and

\begin{equation}
x = 0.85 \bigg(\frac{r}{\mathrm{au}}\bigg) \bigg(\frac{\mathrm{M}_*}{1 \mathrm{M}_{\odot}}\bigg)^{-1},
\end{equation}

\noindent
is the scaling from \citet{owen2012}.

The internal photoevaporation process removes mass from the disc following the profile defined in Eq. \ref{eq:masslossprofile}. If a grid cell contains less mass than is prescribed to be removed, this excess is removed in the nearest outer cell. As the cells are traversed inside-out, this takes the form of inside-out disc clearing. 

\subsubsection{Disc dust photoevaporation}

Circumstellar discs are composed of gas and dust. Initially, the
gas:dust ratio is 100:1. This value is derived from the consideration
that the ratio is inherited from the interstellar medium
\citep{bohlin1978}. Grain growth might result in much lower gas:dust
ratios \citep{williams2014}, and the ratio is likely to change during
the lifetime of a disc \citep{manara2020}. Models of dust evolution
and radial drift \citep{birnstiel2010, rosotti2019a} show that the
dust-to-gas mass-ratio $\delta$ decreases with time. We introduce a
simple prescription for the photoevaporation of dust inside circumstellar discs
in the present work.

We follow the prescription of \citet{haworth2018a} to calculate the
mass loss rate of dust entrapped in the photoevaporation wind. This
mass-loss rate is described as:

\begin{equation}
\dot{M}_{dw} = \delta {\dot{M}_{gas} }^{3/2} \left(\frac{{\nu}_{th}}{4 \pi F G \mathrm{M}_* {\rho}_g a_{min}}\right)^{1/2} \mathrm{exp}\left(\frac{-\delta (G \mathrm{M})^{1/2} t}{2{\mathrm{R}_d}^{3/2}}\right).
\end{equation}

\noindent
Here the initial dust-to-gas mass-ratio $\delta = 10^{-2}$,
$\dot{M}_{gas}$ is the gas mass loss rate (determined as explained in
sections \ref{model:extphotoevap} and \ref{model:intphotoevap}),
${\nu}_{th} = \sqrt{8 k_{b} T / (\pi \mu m_{H})}$ is the mean
thermal speed of the gas particles, $F$ is the solid angle subtended
by the disc at the outer edge, ${\rho}_g$ is the grain mass density
($1 g/cm^3$, \citet{facchini2016}), and $a_{min}$ is the
minimum grain size at the disc radius $\mathrm{R}_d$. We assume
$a_{min} = 0.01 \mu\mathrm{m}$ \citep{haworth2018a, facchini2016}.

This model takes into account, the fraction of the dust entrained in the photoevaporation wind, and how it decreases over time due to dust growth. Mass is removed from a single scalar reservoir. The radial structure is implicitly assumed to follow the gas structure, multiplied by the dust-to-gas ratio $\delta \sim 0.01$, and does not account for the dust fraction enhancement due to the evaporation-resistant dust population.

This reservoir can also be depleted by other processes such as inward
radial drift and the formation of pebbles, planetesimals, and
planets. For this reason, we consider the dust reservoir in this model
to be an upper limit for the total mass in solids with an unknown size
distribution.

\subsubsection{Dynamical truncations}\label{model:truncations}

Circumstellar discs can be truncated in encounters with other stars in the cluster.
We calculate a semi-analytical truncation radius based on \citet{adams2010}, who propose that the new radius of a disc after a truncating encounter is $R' \approx b/3$ where $b$ is the pericentre distance of the encounter. We combine this with the mass dependence of \citet{breslau2014} to define a truncation radius:

\begin{equation}
R' = \frac{r_{enc}}{3}\left(\frac{m_1}{m_2}\right)^{0.32}.
\end{equation}
\noindent
Here $m_1$ and $m_2$ are the masses of the encountering stars. We
ignore the disc orientation, and the equation for truncation radius is
the average truncation radius over all inclinations. We follow
\cite{portegieszwart2016} in defining an initial collisional radius of
$r_{col} = \SI{0.02}{pc}$ for all stars. This value is updated to
$r_{col} = 0.5 r_{enc}$ after every encounter, to guarantee that each
encounter is only detected once within the time step. Not all
encounters result in disc truncation. If the calculated truncation
radius $R'$ exceeds the current radius of an encountering disc, the
disc is not affected by the encounter. If a disc is truncated in an
encounter, we set the new radius of a disc to $R'$ by making the
column density $\Sigma_{\mathrm{edge}} = 10^{-12}\,g/cm^3$ for every
disc cell outside $R'$. The truncated disc then continues to expand
viscously.

Dynamical encounters not only change the disc sizes and strip mass
from the outskirts but can also lead to changes in the mass
distribution of the discs, and mass exchange can occur between the
encountering discs \citep{pfalzner2005, rosotti2014, jilkova2015,
  portegieszwart2016}. We do not consider mass exchange or changes to
the mass distribution during dynamical encounters, other than
truncation. When a disc is truncated in our model, all the mass
outside its new radius $R'$ is simply lost.

\subsection{Initial conditions}

\subsubsection{Molecular cloud}

Our simulations start with a spherical cloud model of mass $10^4$
\MSun and initial radius \SI{3}{pc}. We use 32.000 SPH particles,
which results in a resolution of 0.3 \MSun per particle. The softening
in the simulations is \SI{0.05}{pc}. We use a power-law velocity
spectrum to model large-scale turbulence \citep{bate2003}.  Each
realization of a molecular cloud has a different random seed, and as a
result, the substructure is different in every run. We run 6
realizations of the molecular cloud collapse simulations. These
realizations differ only in the random seed used to determine the
position and velocities of the SPH particles.

\subsubsection{Circumstellar discs}

We choose the initial disc radii as:

\begin{equation}\label{eq:discradius}
r_d(t=0) = R'\left(\frac{M_*}{M_\odot}\right)^{0.5}.
\end{equation}

\noindent
Here $R'$ is a constant. We choose $R' = \SI{30}{au}$, which results in initial disc radii between $\sim\SI{5}{au}$ and $\sim\SI{40}{au}$. This is in agreement with observations that suggest that young circumstellar discs are generally quite compact (radii around 20 to $\SI{50}{au}$, \citet{trapman2020, tobin2020}). 

The initial mass of the discs:

\begin{equation}\label{eq:discmass}
\mathrm{M}_d(t=0) = 0.1 \mathrm{M}_*. 
\end{equation}

\noindent
This yields initial disc masses ranging from $\sim 8 \mathrm{\ M}_{Jup}$ to $\sim 200 \mathrm{\ M}_{Jup}$.

\section{Results}

\subsection{Star formation and cluster evolution}

\begin{figure*}
\includegraphics[width=\linewidth]{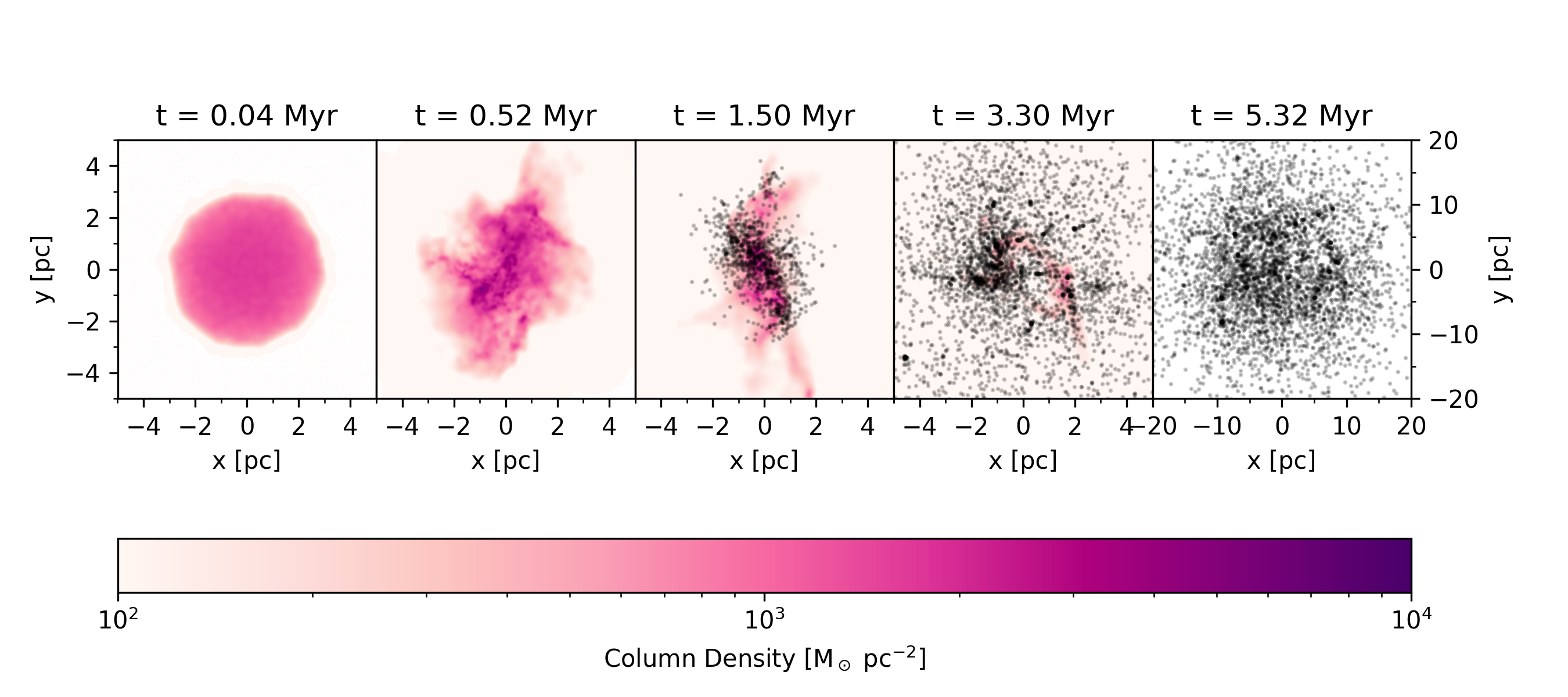}
\caption{The evolution of the molecular cloud collapse and star formation process in run \#4. The second panel shows the moment before the first stars form, and the center panel shows the region during star formation. The fourth panel shows the moment just before gas expulsion, and the rightmost panel shows the region close to the end of the simulation.}
\label{fig:cloud}
\end{figure*}

\begin{figure}
\includegraphics[width=\linewidth]{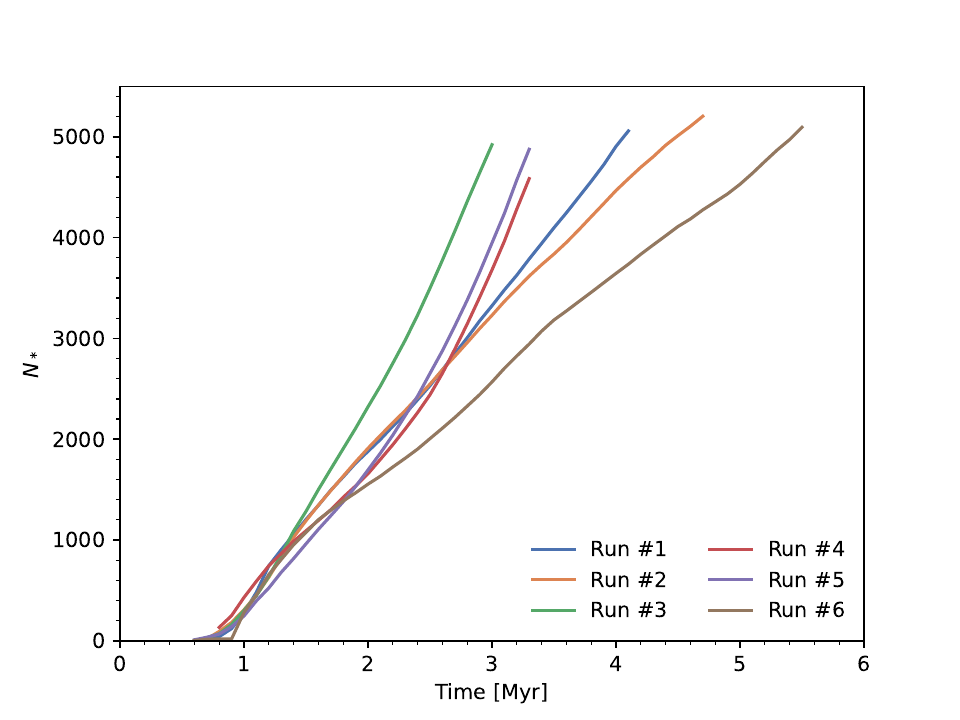}
\caption{Number of stars in time for each simulation run.}
\label{fig:N_vs_t}
\end{figure}

In Figure \ref{fig:cloud} we illustrate the evolution of the molecular cloud collapse and star formation process in one of our simulations. In Figure \ref{fig:N_vs_t} we show the number of stars in time for each simulations. In Table \ref{table:dyn} we present the final number of stars in each run. The mean number of stars created in six runs is $5031 \pm 198$.

\begin{figure}
\includegraphics[width=\linewidth]{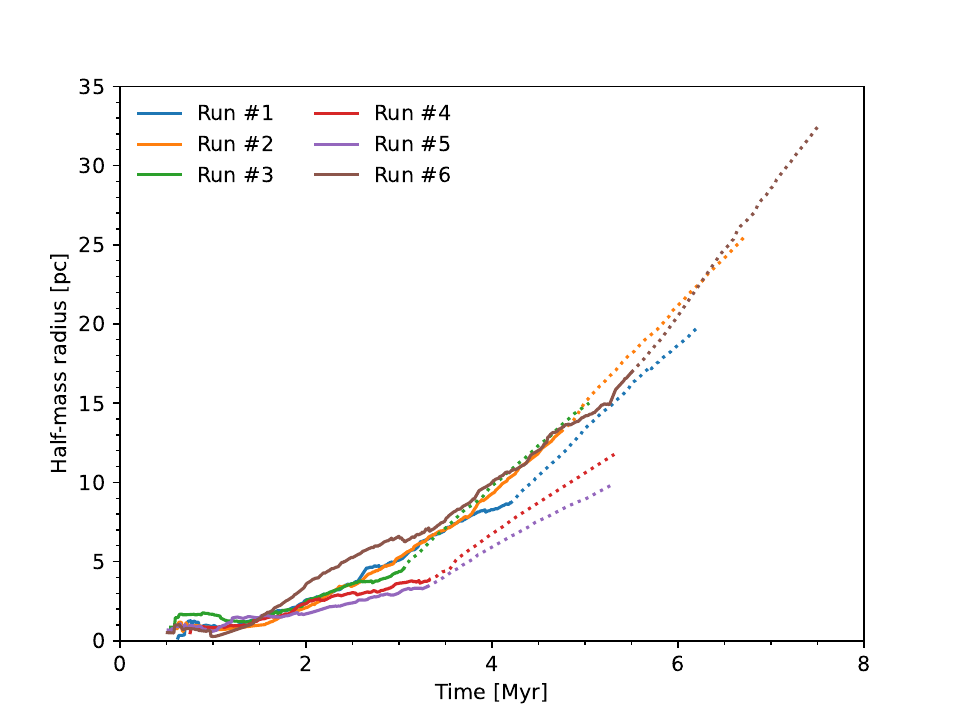}
\caption{Cluster half-mass radius of the simulations in time. The solid lines
  correspond to the half-mass radius while star formation is still
  ongoing and the dotted lines after it has ended.}
\label{fig:Rvir}
\end{figure}

In Figure \ref{fig:Rvir} we show the evolution of the cluster half-mass radius
in time for each of our simulations.  The solid lines show the
half-mass radius while star formation is still ongoing, whereas the
dotted lines follow the radius after all stars have formed and gas has
been removed from the clusters. All regions initially expand gradually
at a rate of 1 to 2~pc/Myr. After gas expulsion this expansion
accelerates to a rate of 3 to 10~pc/Myr.

To quantify the spatial distribution of the stars resulting from the
molecular cloud collapse simulations, we look at the Q parameter of
the minimum spanning tree \citep{cartwright2004} and the fractal
dimension in each region at the end of star formation. The Q parameter
is calculated as:

\begin{equation}
Q = \frac{\overline{m}}{\overline{s}}.
\end{equation}

\noindent
Here $\overline{m}$ is the mean length of the minimum spanning tree, and $\overline{s}$ is the mean separation between the stars. Regions with $\mathrm{Q} > 0.8$ are smooth and centrally concentrated, while values of $\mathrm{Q} < 0.8$ correspond to regions with substructure.

\begin{figure}
\includegraphics[width=\linewidth]{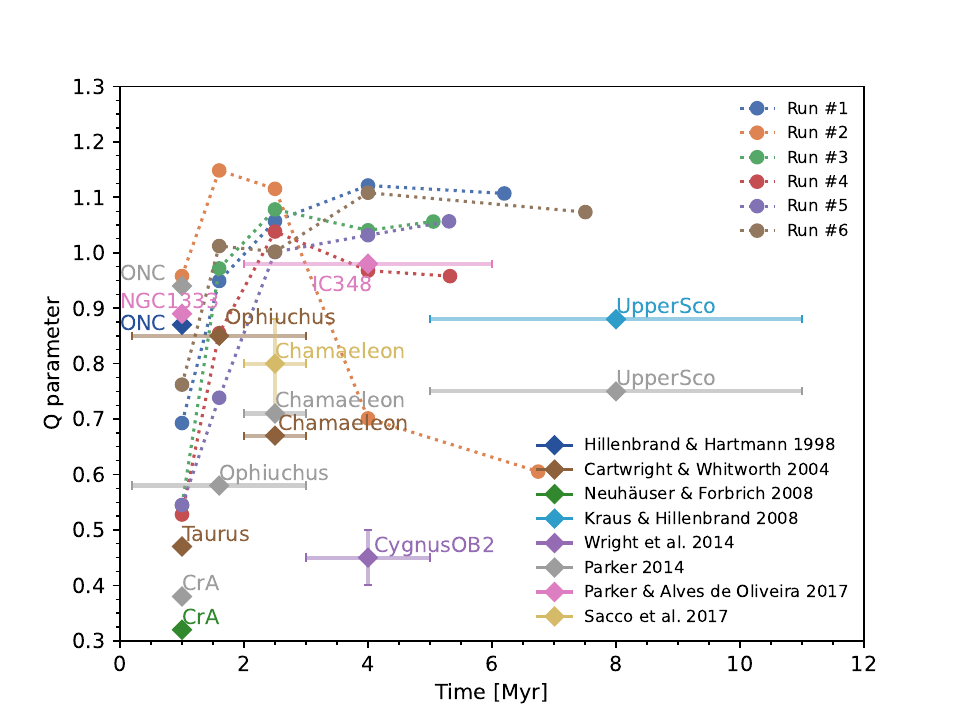}
\caption{Q parameter of our simulations in time, and values for observed star-forming regions. The Q parameter in our simulations considers only stars with masses $M_* > 0.5 \mathrm{\ M}_{\odot}$. The Q parameters are computed at times corresponding to the ages of the observed star-forming regions and the end of the simulation, and shown as points connected by linear dashed lines to guide the eye. }
\label{fig:q}
\end{figure}

In Figure \ref{fig:q} we show the Q parameter at time for each of our
simulations, along with values for several observed regions. The Q
parameter of the simulations is calculated from a 2D projection of the
stellar distances, and considers only stars with masses $M_* > 0.5
\mathrm{\ M}_{\odot}$. In Table \ref{table:obs} we summarize the
values for Q and the estimated ages for each region, along with the
corresponding references. The simulation results span a range of
$\mathrm{Q} \sim 0.5 - 0.8$ when star formation starts and evolve to
$\mathrm{Q} \sim 0.9 - 1.1$.

Run \#2 forms a bit of an exception is several regards.  Part of its
deviating values (see Tab.\,\ref{table:dyn}), in $\mathrm{Q}$ but also
in the fractal dimension probably result from an event that occurred
some time during the growth of two massive sinks.  These two sinks are
ejected from the cluster at a velocity of 200 and 600 km/s, possibly
resulting from a three-body interaction. By the end of the simulation
both sinks have produced $\sim 100$ stars that co-move with their
parent sinks.  We did not want to remove this run from the
simulations, even though we suspect that this ejection is the result
of a numerical error in the hydrodynamics solver. We have to note
there that the combined contributions of all the processes in our
simulations make it hard to check for energy conservation, because in
the combination of irradiation, internal disc evolution, stellar
evolution and gravitational dynamics such conservation laws do not
apply locally.  The main cluster, from which the two smaller clumps
are ejected, turns out rather usual, and we decided to take this
cluster into account in the further analysis.

In observations, the Q parameter can greatly vary depending on stellar membership. Regions with lower Q, such as Corona Australis \citep{parker2014b}, Cygnus OB2 \citep{wright2014}, and Taurus \citep{cartwright2004} are highly substructured. In observations of star-forming regions, Q might vary depending on membership uncertainty \citep{parker2012}. This leads to regions with more than one Q value, such as Corona Australis, Chamaeleon, the ONC, Ophiuchus, and Upper Scorpio. 

As can be seen in Figure \ref{fig:q}, the final shapes of the clusters generated by our simulations are smoother than those of observed clusters. The smootheness of our simulated clusters might be caused by the absence of stellar feedback \citep[as was also argued in][see also sect.\,\ref{sec:discussion}]{maclow2004, hansen2012, offner2015}. 

\begin{table}\caption{Run number, final number of stars ($N_*$), time of the end of star formation ($\mathrm{t}^{\mathrm{SF}}_{\mathrm{end}}$), Q parameter, and fractal dimension ($F_d$) for our simulation results.}
\centering
{%
\begin{tabular}{@{}lccll@{}}
\toprule
Region & $N_*$ & $\mathrm{t}^{\mathrm{SF}}_{\mathrm{end}}$ {[}Myr{]} & $\mathrm{Q}_{\mathrm{end}}$ & $F_d$ \\ \midrule
Run \#1 & 5198 & 4.20 & 1.11 & 1.8 \\
Run \#2 & 5242 & 4.75 & 0.61 & 1.0 \\
Run \#3 & 5071 & 3.05 & 1.06 & 1.8 \\
Run \#4 & 4654 & 3.32 & 0.96 & 1.6 \\
Run \#5 & 4914 & 3.31 & 1.06 & 1.6 \\
Run \#6 & 5106 & 5.51 & 1.07 & 1.7 \\
\bottomrule
\end{tabular}%
}\label{table:dyn}
\end{table}

\begin{table*}\caption{Region name, number of stars ($N_*$), age, Q parameter, and fractal dimension ($F\_d$) for our simulation results and observed regions. References: (a) \citet{parker2014b}; (b) \citet{neuhauser2008}; (c) \citet{luhman2016}; (d) \citet{parker2017}; (e) \citet{hillenbrand1998}; (f) \citet{cartwright2004}; (g) \citet{hartmann2002}; (h) \citet{kraus2008}; (i) \citet{simon1997}; (j) \citet{bontemps2001}; (k) \citet{luhman2007}; (l) \citet{wright2010}; (m) \citet{wright2014}; (n) \citet{carpenter2006}; (o) \citet{luhman2012}; (p) \citet{sacco2017}; (q) \citet{galli2020}; (r) \citet{luhman2020}; (s) \citet{luhman2018}.}
\resizebox{\textwidth}{!}{%
\begin{tabular}{@{}lccll@{}}
\toprule
Region & $N_*$ & Age {[}Myr{]} & Q & $F_d$ \\ \midrule
Corona Australis (CrA) & $\sim 313^{(q)}$ & $\sim 1.0^{(a)}$ & $0.32^{(b)}$, $0.38^{(a)}$ & - \\
NGC 1333 & $\sim 200^{(c)}$ & $\sim 1.0^{(c)}$ & $0.89^{(d)}$ & - \\
ONC & $\sim 1000^{(e)}$ & $\sim 1.0^{(e)}$ & $0.87^{(e)}$, $0.94^{(a)}$ & - \\
Taurus & $\sim 438^{(s)}$ & $\sim 1.0^{(f)}$ & $0.47^{(f)}$ & $1.5^{(f)}$, $1.02 \pm 0.04^{(g)}$, $1.049 \pm 0.007^{(h)}$, $1.5 \pm 0.2^{(i)}$ \\
Trapezium & $\sim 1000^{(e)}$ & $\sim 1.0^{(e)}$ & - & $1.5 \pm 0.2^{(i)}$ \\
Ophiuchus & $199^{(f)}$ & $1.6 \pm 1.4^{(j)}$ & $0.85^{(f)}$, $0.58^{(a)}$ & $1.5 \pm 0.2^{(i)}$ \\
Chamaeleon I & $120^{(p)}$ & $2.5 \pm 0.5^{(k)}$ & $0.67^{(f)}$, $0.71^{(a)}$, $0.80 \pm 0.08^{(p)}$ & $2.25^{(f)}$ \\
Cygnus OB2 & $\sim 2700^{(l)}$ & $4.0 \pm 1.0^{(l)}$ & $0.45 \pm 0.05^{(m)}$ & - \\
IC 348 & $\sim 500^{(c)}$ & $4.0 \pm 2.0^{(c)}$ & $0.98^{(d)}$ & - \\
Upper Scorpio & $\sim 1761^{(r)}$ & $8.0 \pm 3.0^{(n)}$ & $0.88^{(h)}$, $0.75^{(a)}$ & $0.69 \pm 0.09^{(h)}$ \\ \bottomrule
\end{tabular}%
}\label{table:obs}
\end{table*}

\begin{figure}
\includegraphics[width=\linewidth]{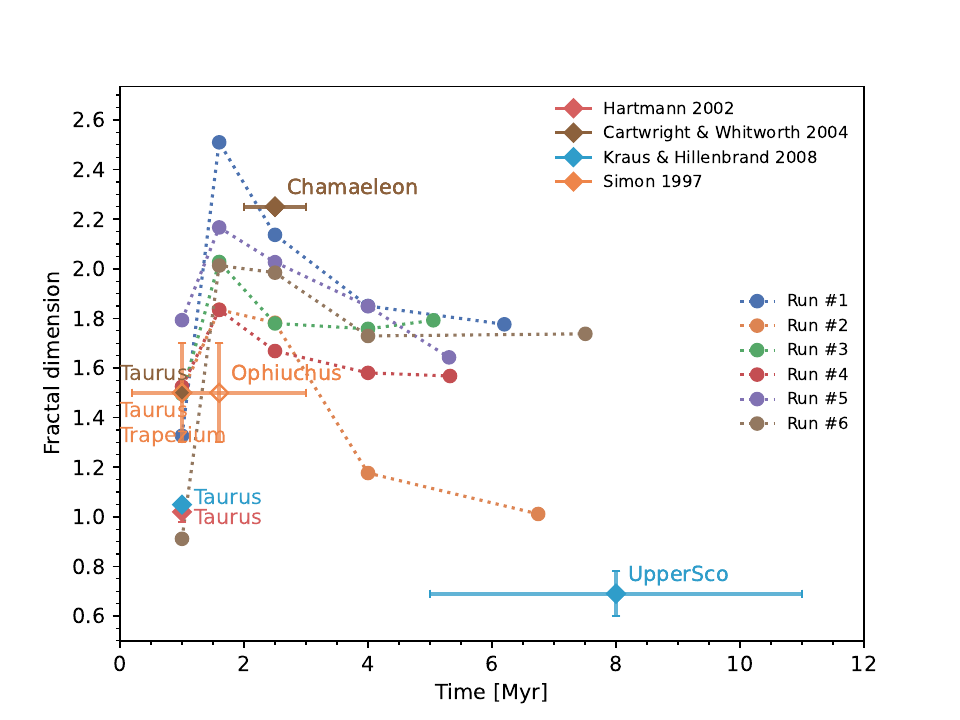}
\caption{Fractal dimension of our simulations in time, and values for observed star-forming regions. The fractal dimension in our simulations considers only stars with masses $M_* > 0.5 \mathrm{\ M}_{\odot}$. The fractal dimensions are computed at times corresponding to the ages of the observed star-forming regions and the end of the simulation, and shown as points connected by linear dashed lines to guide the eye.}
\label{fig:fdtime}
\end{figure}

Another way to quantify the structure of a star-forming region is by measuring its fractal dimension, $F_d$. The fractal dimension is a measurement of the clumpiness of a region. Low values of $F_d$ ($\lesssim 1.5$) signify regions with important substructure. Higher values ($\sim 2.0$ to 3.0) mean that the regions are smoother \citep{goodwin2004, delafuentemarcos2006}. In Figure \ref{fig:fdtime} we show the evolution of the fractal dimension in time for each simulation. Initially they span a range from $F_d \sim 0.9-1.8$, and first increase before decreasing to a general range of $F_d \sim 1.5-1.8$. Again, run \#2 is an exception due to its runaway clusters. 

\subsection{Disc masses}\label{results:discs}

\begin{figure*}
\includegraphics[width=\linewidth]{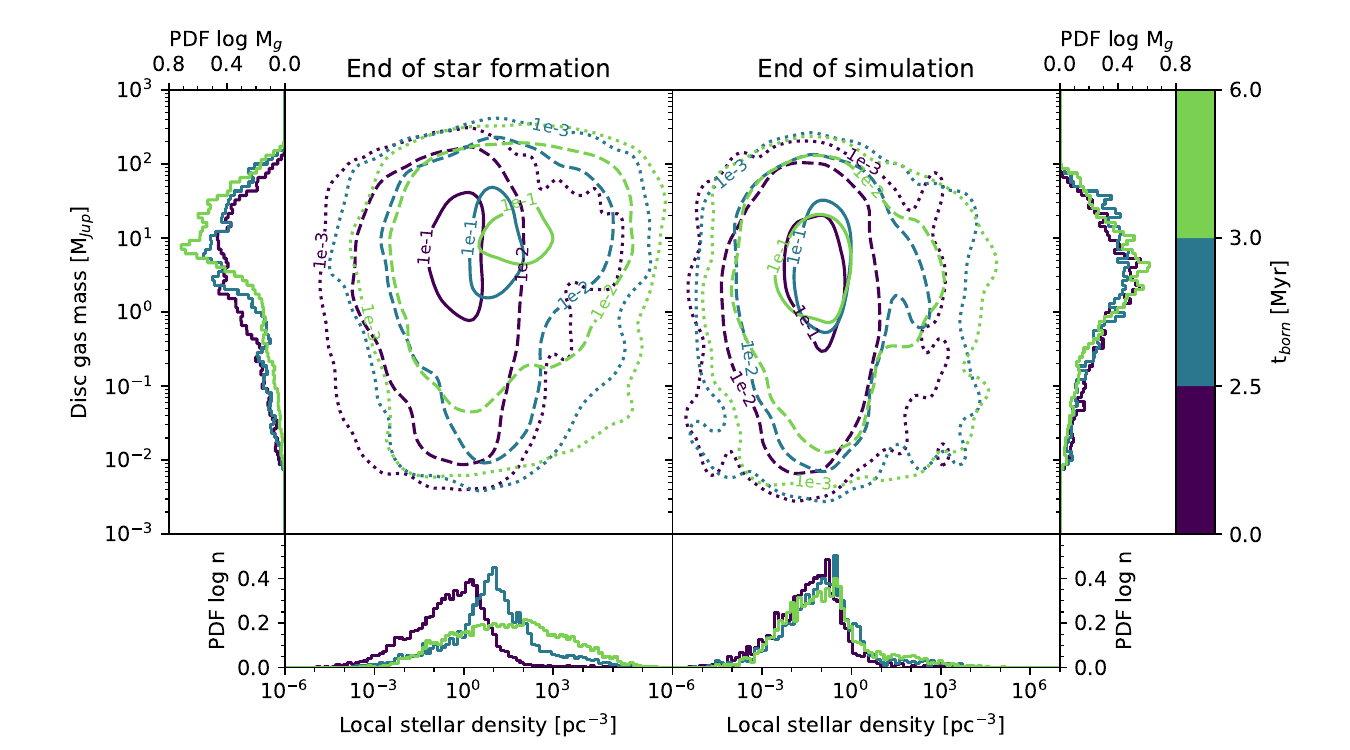}
\caption{The distribution of local stellar density and gas mass of
  three disc populations, at two moments in the simulation. The left
  panels show the population at the end of star formation, the right
  panels at the end of the simulation. The populations are: discs born
  between \SI{0}{Myr} and \SI{2.5}{Myr} (purple, 6814 and 1645 discs
  at the end of star formation and the simulation, respectively);
  discs born between \SI{2.5}{Myr} and \SI{3}{Myr} (blue, 4465 and
  1410 discs); and discs born between \SI{3}{Myr} and \SI{6}{Myr}
  (green, 6777 and 2754 discs).}
\label{fig:contours_timeborn}
\end{figure*}

\begin{figure*}
\includegraphics[width=\linewidth]{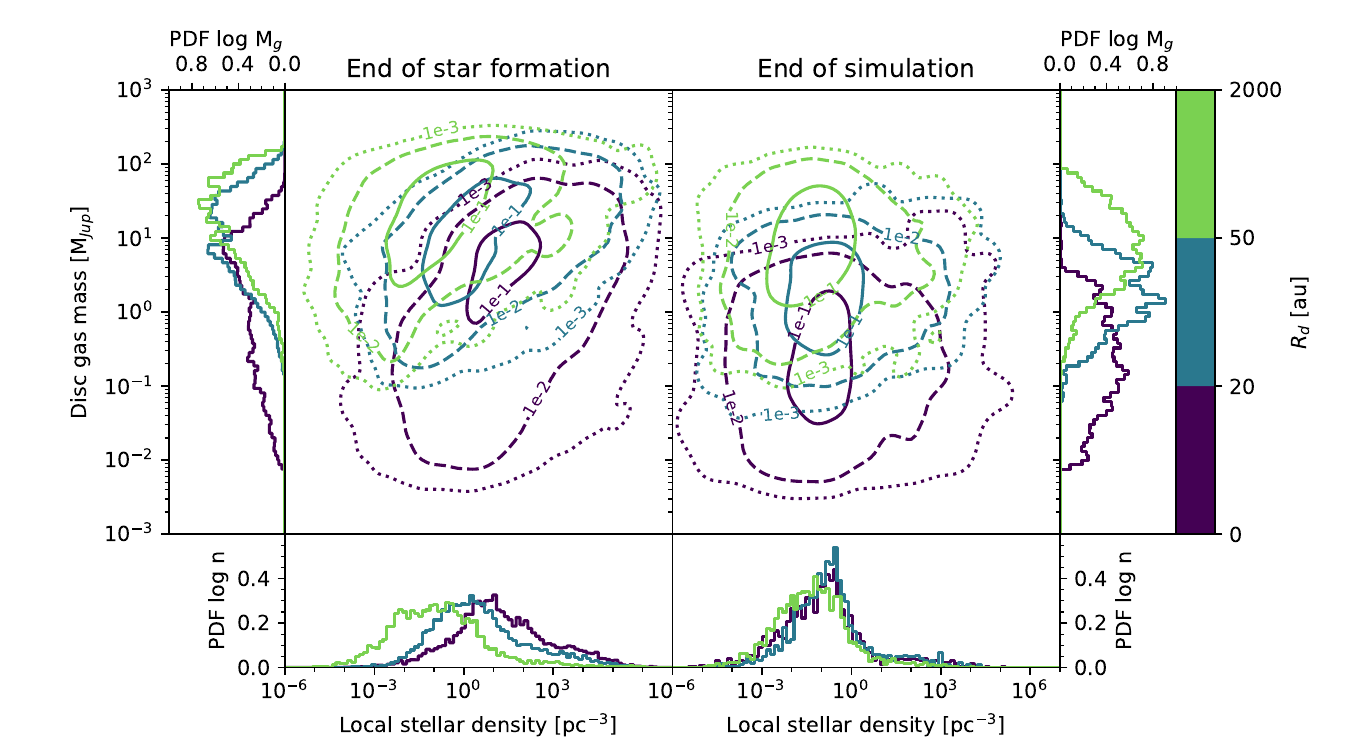}
\caption{The distribution of local stellar density and gas mass of
  three disc populations, at two moments in the simulation. The left
  panels show the population at the end of star formation, the right
  panels at the end of the simulation. The populations are: discs with
  radii between \SI{0}{au} and \SI{20}{au} (purple, 8216 and 1437
  discs at the end of star formation and the simulation,
  respectively); discs with radii between \SI{20}{au} and \SI{50}{au}
  (blue, 6366 and 1260 discs); and discs with radii between
  \SI{50}{au} and \SI{2000}{au} (green, 3474 and 3112 discs).}
\label{fig:contours_radius}
\end{figure*}

In Figures \ref{fig:contours_timeborn} and \ref{fig:contours_radius}
we show the distribution of disc mass versus local number density
for the non-dispersed discs in all simulation runs. The left panels
show the discs at the end of the star formation process. The right
panels show the discs at the end of the simulations, \SI{2}{Myr} after
star formation has finished. The 2D distributions shown are the
contours of the probability density, estimated using Gaussian kernel
density on the logarithm of disc mass and stellar density. The 1D
distributions are estimated using histograms of the logarithm of the
respective quantity.

Furthermore, we separate the discs in three populations, based moment
of their birth (relative to the start of the simulation) in Figure
\ref{fig:contours_timeborn} and their disc radius in Figure
\ref{fig:contours_radius}. These bins were chosen to have comparable
numbers of discs in all bins at both times.

We calculate the local stellar density using the method by
\citet{casertano1985} with the five nearest neighbors. While star
formation is still ongoing, stars and discs form in regions spanning
the whole range of stellar density. In particular, given our
implementation of the star formation process with sink particles, many
stars tend to form in regions of high stellar densities. Once the star
formation process ends and the gas is expelled, the clusters expand.
This results in a decrease in the overall density.  The consequence is
that discs are less often harassed dynamicall, but also that
photoevaporation becomes less effective.  By the end of the
simulations, almost no discs are present in regions of local density
$\gtrsim 10^5$ stars $\mathrm{pc}^{-3}$.

\clearpage

In the bottom panels of Figure \ref{fig:contours_timeborn} we can see
how the local stellar density evolves. At the end of star formation,
the stars born late (3-6 Myr) have not yet had time to evolve far from
the gas distribution they formed in and form a rather symmetric,
almost log-normal density distribution, similar to that of turbulent
gas \citep{krumholz2014}. The older populations form a less symmetric
distribution, with a heavier tail towards low densities. At the end of
the simulation, each population has virialized and the density
distributions of each of the populations are similar.

A similar trend is visible in the local stellar density of the disc
radius bins. At the end of star formation, large discs are found in
low-density regions, and small discs in high-density regions. 2 Myr
later, the distributions of the three bins are similar.

At the end of star formation, the disc mass distributions of the three
age bins peak at very similar values, but are subtly different at
extreme values. The greatest disc masses belong to young discs, and
smaller disc masses belong mostly to older discs. However, at the end
of the simulation the distributions become more similar. The number of
discs in each population differs considerably: of the old population
about a quarter of discs survive, while of the young population about
40\% survives.

There is a clear trend for more massive discs to have larger
radii. This is not simply a result of the initial conditions and the
viscous expansion of the discs.

At the start of the simulations, the largest disc had a radius of 41
au. All discs that exceed this value at a later time have viscously
grown to that size. External photoevaporation (and truncation) causes
the discs to shrink, because we remove the evaporated gas from the
outside.  This process opposing viscous growth.  At the end of our
simulations the largest discs tend to be those that lose little mass
and, therefore, are also among the most massive.

\begin{figure}
\includegraphics[width=\linewidth]{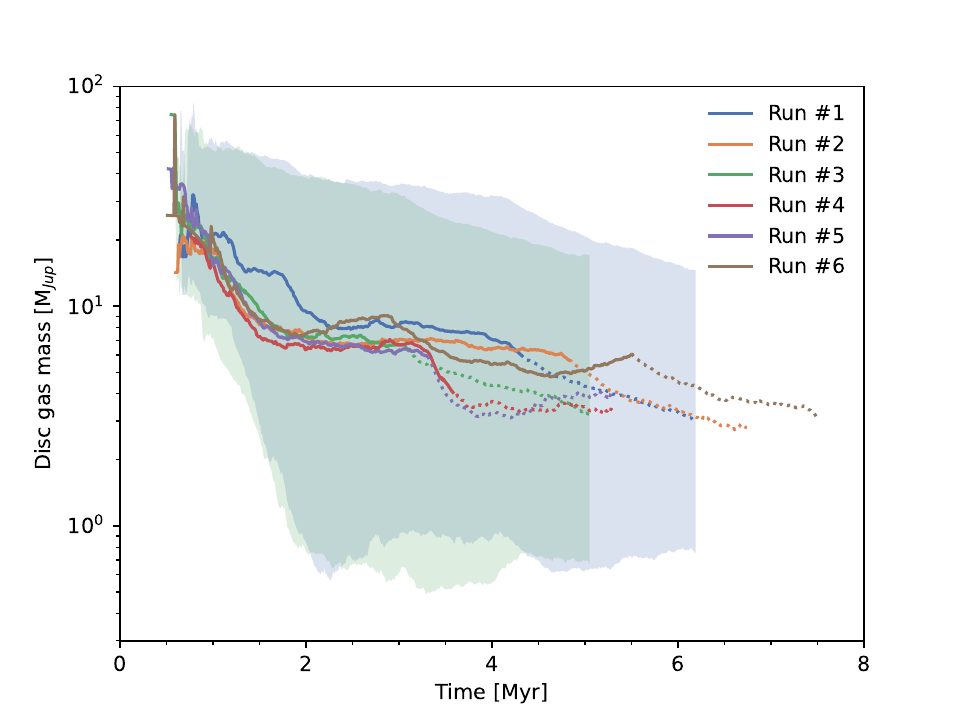}
\caption{Median disc mass in time for each simulation run. The solid lines correspond to ongoing star formation, and the dotted lines after it has ended. The shaded areas span the range between the \nth{16} percentile and the \nth{84} percentile. For clarity, these ranges are shown only for 2 runs, but the other ones span similar magnitudes.}
\label{fig:meanmass_vs_time}
\end{figure}

In Figure \ref{fig:meanmass_vs_time} we present the median disc mass as a function of time. The solid lines show the times while star formation is still ongoing, and the dotted lines once the last star has formed. The mean disc mass varies during star formation because
new discs are still forming and directly exposed to external photoevaporation. Once star formation stops, discs also stop forming, and, as a result, the median disc mass generally drops. In run \#4, the median disc mass remains relatively constant, and in run \#5, it even increases. This is possible if low-mass discs are preferentially dispersed while high-mass discs lose little mass. 

\begin{figure}
\includegraphics[width=\linewidth]{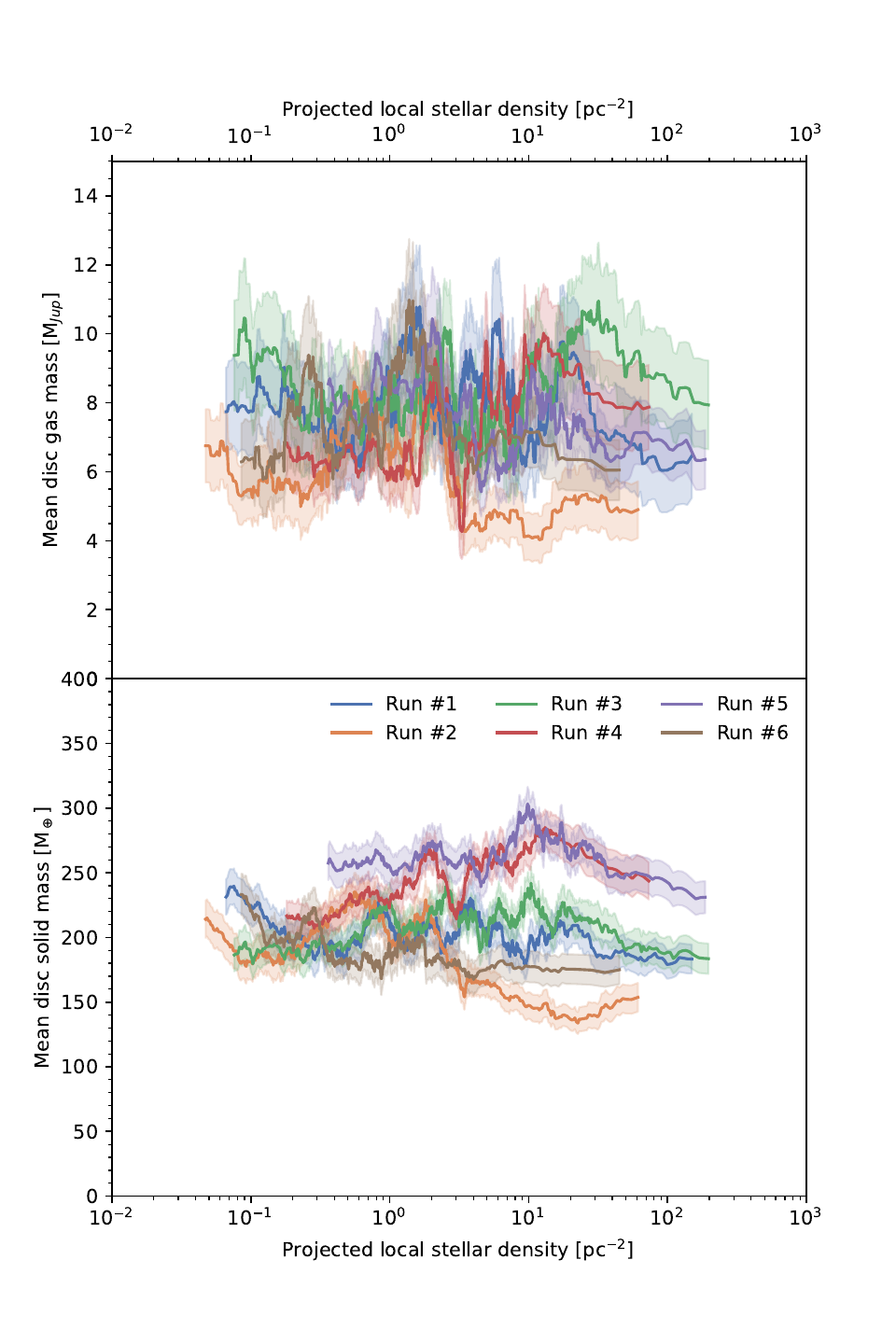}
\caption{Binned mean gas mass (top panel, $\mathrm{M}_{Jup}$) and solid mass (bottom panel, \MEarth) at the end of the simulations versus projected local stellar density. The binned mean is calculated using a rolling bin spanning 100 stars. The local stellar number density is calculated as specified in section \ref{results:discs}.}
\label{fig:binned}
\end{figure}

In Figure \ref{fig:binned} we present the binned mean gas and solid
masses of the discs as a function of the projected local stellar
number density at the end of the simulations. The local stellar
density is calculated in the same way as for Figure
\ref{fig:contours_timeborn}, but with the distances between stars
projected to two dimensions. The binned mean is calculated using a
rolling bin spanning 100 stars.  Both gas mass and solid mass are
relatively constant across densities.

\begin{figure}
\includegraphics[width=\linewidth]{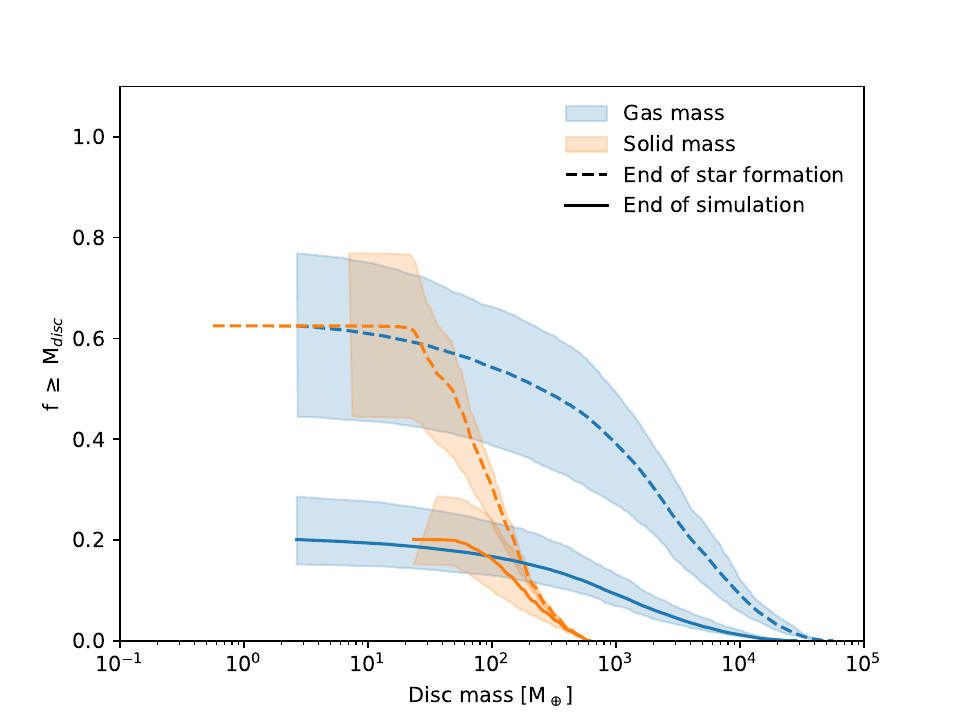}
\caption{Cumulative distribution of disc solid and gas masses at the end of star formation (dashed lines) and at the end of the simulations (solid lines). The lines show the mean values for all runs and the shaded areas show the standard deviation.}
\label{fig:cumulativemass}
\end{figure}

In Figure \ref{fig:cumulativemass} we show the cumulative distribution
of disc gas and solid masses at the end of star formation and the end
of the simulations. Note that we normalize to all stars that have
hosted discs. If the cumulative distribution has a maximum of 0.6,
that means 60\% of stars retain discs.  The lines show the cumulative
distribution of discs from all our simulations. The shaded regions are
between the extremes of the distributions of the individual
simulations.  By the end of star formation, almost all remaining discs
have masses in solids $\gtrsim 20 \mathrm{M}_{\oplus}$, up to $\sim
600 \mathrm{M}_{\oplus}$. After \SI{2}{Myr} of evolution, the
distribution spans almost the same range. All of the final discs have
mass insolids in excess of $10 \mathrm{\ M}_{\oplus}$, which is a
lower limit mass for the formation of rocky planets and the cores of
gas giants \citep{ansdell2016}. The masses obtained in the present
simulations are higher than those in \citet{concha-ramirez2021},
suggesting that the extended period of star formation may mediate the
survival of massive discs. We expand on this discussion in Section
\ref{sec:discussion}.

\section{Discussion}\label{sec:discussion}

\subsection{Comparison with previous work}

In previous work \citep{concha-ramirez2019a, concha-ramirez2021} we
performed simulations of star clusters in which stars with masses
$\mathrm{M}_* \leq 1.9 \mathrm{M}_{\odot}$ have circumstellar
discs. These discs were subject to viscous evolution, external
photoevaporation, and dynamical truncation. We assumed discs to be
composed of 1\% dust (by mass). In those previous simulations, we
found that discs become depleted of mass sufficiently quickly that
$\sim 60\%$ to 90\% are dispersed within \SI{2}{Myr}. In these
simulations, all stars were born instantaneously in a virialized
Plummer distribution.

Here, we improve these models in two main ways: First, we
start with a giant molecular cloud that forms stars through hydrodynamical collapse. This improved treatment of cluster formation leads to non-spherical and non-equilibrium initial stellar distributions.
It also leads to stars being born over a time frame of 2\, to 5\,Myr.
A second improvement is to the circumstellar disc model. We implement a model for external photoevaporation to remove the discs' dust \citep{haworth2018a} and for internal photoevaporation to remove the discs' gas \citep{owen2012,picogna2019}.

These improvements to the initial model by \citet{concha-ramirez2019a, concha-ramirez2021} resulted in interesting differences in the results. During the first few million years of evolution, the star formation process is ongoing and, as discs lose mass due to photoevaporation, the new discs that are constantly being formed keep the median mass of the overall population relatively constant (Figure \ref{fig:meanmass_vs_time}). As soon as this constant replenishing of discs stops, in most runs the median mass of the discs quickly drops (we discuss the runs with a different trend below). This is consistent with the results from \citet{concha-ramirez2019a} and \citet{concha-ramirez2021}. However, the mass in solids in our simulations (see the bottom panel of Figure \ref{fig:binned}) is much higher than in \citet{concha-ramirez2021} in regions of comparable stellar projected density. 

This discrepancy is partly due to a difference in IMF; our previous
work used a lower limit of 0.01 M$_\odot$ (which includes brown
dwarfs), while here we use a lower limit of 0.08 M$_\odot$ (the
hydrogen burning limit). This explains why the average final solid
masses in this work are even higher than the average initial dust
masses in our previous work. However, the decrease in dust
photoevaporation also plays a role, essentially locking in the solid
content after $\sim$1 Myr. In \citet{concha-ramirez2021}, the
dust-to-gas ratio was assumed to be constant throughout the
simulation.



\subsection{Star formation recipe}\label{sec:sfr}

The spatial distribution of star-forming regions also impacts
disc-mass distributions. In previous work we performed simulations
starting from a spherically symmetric star cluster. Here we model the
collapse of a molecular cloud to improve on these initial
conditions. The resulting clusters, however, are, on average, smoother
than observed star-forming regions (see Figure \ref{fig:q}). Stellar
feedback, in particular stellar winds and jets from protostars, can
substantially affect the morphology of star-forming regions
\citep[e.g.][]{maclow2004, hansen2012, offner2015}.

Moreover, the SFE in our model is overestimated. While we set up an SFE of 30\% for the star formation process, the integrated SFE of molecular clouds may be orders of magnitude smaller \citep[e.g.][]{chevance2020a}, and efficiencies of the order used in this work are found within much smaller clumps \citep[$\sim 0.1$ pc radius, see e.g.][]{matzner2000}. The star formation efficiency per free fall time is also overestimated. The free fall time of our initial cloud is $\sim$1 Myr, and we form 30\% of the initial gas mass in stars over a few Myr. This implies an efficiency per free fall time of $\sim$10\%. However, measurements of this quantity through various traces indicate a value closer to 1\% \citep{krumholz2019}. 

For this reason we ran two additional series of simulations where we varied the recipe for star formation, but used the same initial conditions and realizations of the IMF. 

In the first series we lowered the absolute star formation efficiency from 0.3 to 0.1. These runs were identical to the runs presented above (up to propagating numerical errors) until they reached the prescribed star formation efficiency, between 1 and 3 Myr before the runs with higher efficiency. We refer to these runs as the low-efficiency runs. 

\begin{figure}
\includegraphics[width=\linewidth]{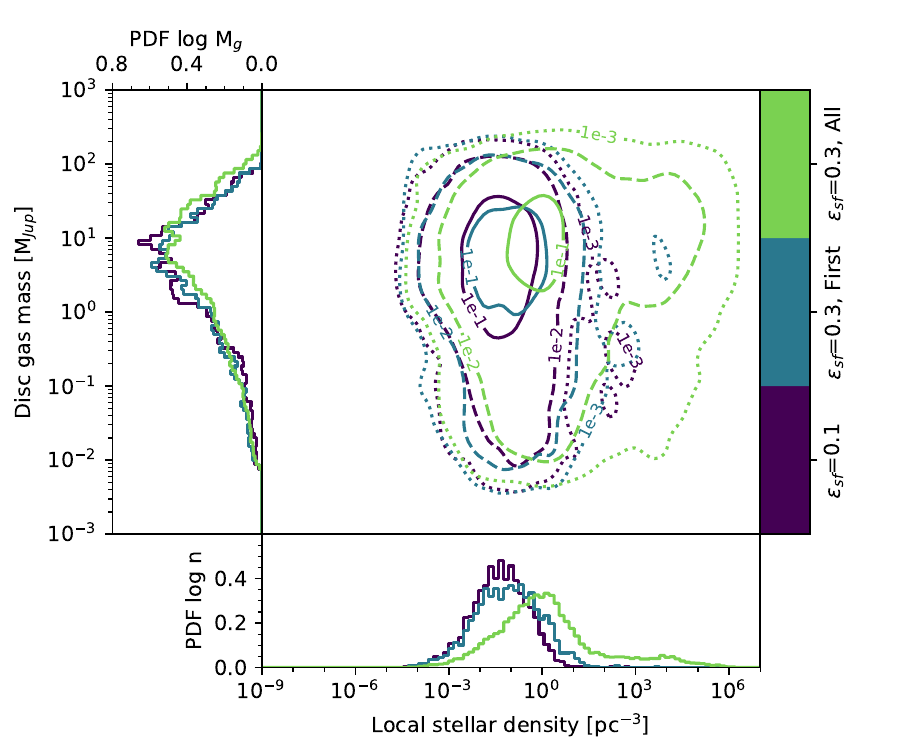}
\caption{The distribution of local stellar density and gas mass of three disc populations. The populations are: all discs in the low-efficiency runs (purple, 4064 discs); the discs in the high-efficiency runs born early (blue, 3598 discs); and all discs in the high-efficiency runs (green, 15130 discs). The data from the low-efficiency runs are from the end of the runs. The data from the high-efficiency runs are from the moment that the low-efficiency run with the same initial conditions ends. }
\label{fig:contours_efficiency}
\end{figure}

We present the distribution of the discs' local stellar density and gas mass in Figure \ref{fig:contours_efficiency}. We compare three populations at the same moment in time, namely the end of the low-efficiency runs, 2 Myr after the end of the star formation process. The three populations are the discs in the low-efficiency runs, the discs in the high-efficiency runs born before the moment star formation ended in the low-efficiency runs (effectively the equivalent of the low-efficiency runs' populations), and the all discs in the high-efficiency runs (born until that moment; star formation has not necessarily ended in these runs).

Comparing the full populations of both sets of runs, we can see that
the disc mass distributions peak around the same mass ($\sim$10
M$_\mathrm{Jup}$), but that of the high-efficiency runs have more
discs at higher disc masses, while the low-efficiency runs have more
discs slightly below the peak. The low mass tails are
indistinguishable. Between the population of the low-efficiency runs
and the first discs of the high-efficiency runs, the disc-mass
distribution of the low-efficiency populations peaks at slightly
higher masses than the early high-efficiency populations.  Two effects
compete here: The first is the late formation of discs in the high-efficiency run,
which leads to the presence of more massive discs (corresponding to those formed
recently). The second is the formation of a larger number of massive stars in
the high-efficiency runs. This leads to more efficient external photoevaporation
of the early population of the high-efficiency runs. This is visible in the slight
shift of the peak of the disc mass distribution.

In the second series of runs we also lowered the rate of star
formation, by increasing the parameter $t_d$ in Equation
\ref{eq:delay} to 10 Myr. We kept the star formation efficiency at
0.1. These runs differ from the main runs presented from the moment
stars form. We refer to these runs as the low-rate runs. In the runs
with a low star-formation rate, the epoch of star formation was a few
Myr longer than in the runs with high star formation. Together with
the decreased absolute SFE, this puts the SFE per freefall time close
to the observed value.

\begin{figure}
\includegraphics[width=\linewidth]{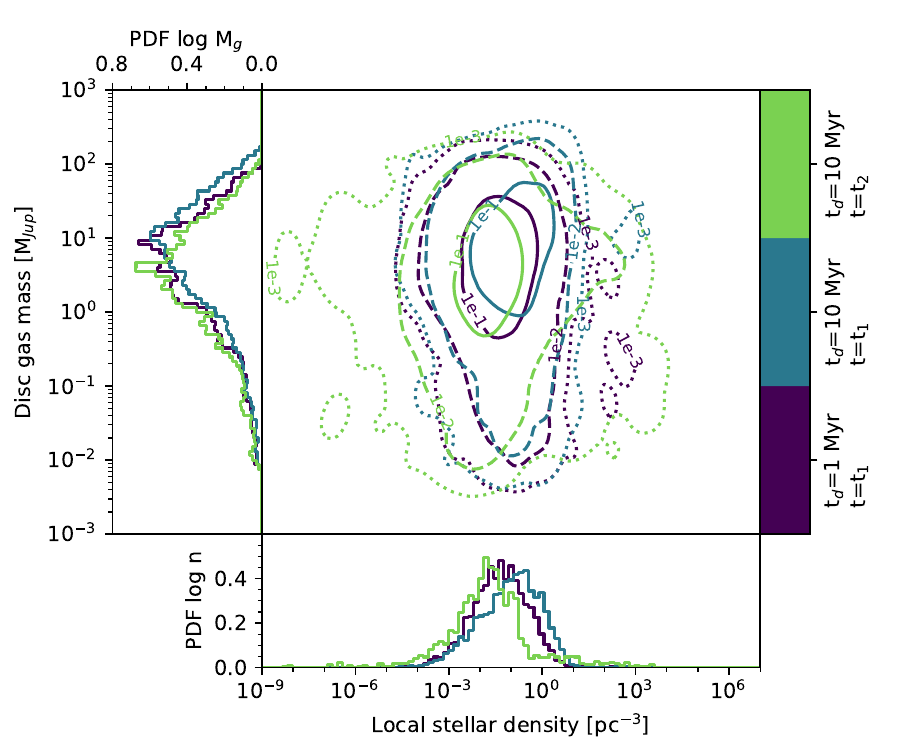}
\caption{The distribution of local stellar density and gas mass of three disc populations. The populations are: discs in the low-efficiency, high-rate runs at the end of those simulations, which we call $t_1$ (purple, 4064 discs); discs in the low-efficiency, low-rate runs at $t_1$ (blue, 4458 discs); and discs in the low-efficiency, low-rate runs at the end of those simulations, which we call $t_2$ (green, 1579 discs).}
\label{fig:contours_rate}
\end{figure}

We present the distribution of the discs' local stellar density and gas mass in Figure \ref{fig:contours_rate}. We compare three populations: the discs in the low-efficiency runs at high and low rates, at the moment the high-rate runs end; and those in the low-effiency, low-rate runs, at the end of those simulations. 

The disc mass distribution of the low-rate runs compared to the high-rate runs at the same moment is shifted to greater masses. At the end of the low-rate runs, the distribution has shifted towards lower masses. Again, it appears that a population with more young discs also has more massive discs.

\subsection{Comparison with observations}

\citet{ansdell2016} propose that a rapid depletion of gas mass in discs can lead to the observed characteristics of exoplanet populations. Traditional theories of planet formation indicate that $\sim 10 \mathrm{M}_{\oplus}$ cores should be able to accrete gaseous envelopes at a rate of $\sim 1 \mathrm{M}_{Jup}$ within $\sim \SI{0.1}{Myr}$. Observational surveys, on the other hand, indicate that ``super-Earths'', or intermediate-mass rocky planets, are about an order of magnitude more common than gas giants \citep[e.g.][]{petigura2013}. \citet{ansdell2016} argue that, if typical $\sim 10 \mathrm{M}_{\oplus}$ cores form in discs that are already depleted of gas, this contributes to the observed diversity in the number of planetary types. Results by \citet{lopez2018} indicate that late ($\gtrsim$ \SI{10}{Myr}) rocky planet formation could explain the newly-observed population of
highly-irradiated rocky planets of $\sim 1.5 \mathrm{R}_{\oplus}$, supporting the idea that rocky planets form in discs which have already lost most of their gas. The evolution of dust and gas disc masses in our simulations suggests a similar context for planet formation. \citet{sellek2020} demonstrate that photoevaporation of dust leads to a decrease in pebble flux in the inner disc after the pebble flux has already peaked, but before 1 Myr. This implies that external photoevaporation of dust need not inhibit the formation of cores.

We do note that our simulations are of massive, high-density star forming regions with massive stars. Locally, the various Orion star forming regions are analogues. However, other regions such as the Lupus and Taurus clouds are of lower mass and density, and lack massive stars. The FUV radiation field in such regions is typically small, and may even be dominated by stars outside the region. \citet{cleeves2016} estimate that one member of the Lupus clouds may be receiving a radiation field of $\sim$3 G$_0$ due to nearby OB stars. \citet{haworth2017} then showed that this radiation field is driving a photoevaporating flow. The photoevaporation rates were computed using a method predating the more carefully modeled FRIED grid. Because the FRIED only provides photoevaporation rates for radiation fields down to 10 G$_0$, we are unable to self-consistently model low-mass star forming regions. 

Circumstellar discs are observed in regions where external radiation should have evaporated them already. 
This `proplyd lifetime problem' is notoriously present in the ONC, where
discs observed within \SI{0.3}{pc} of the massive star $\theta^{1}C$ Ori are exposed to strong radiation fields \citep[$\sim 10^4 \mathrm{\ G}_0$, e.g.][]{odell1994}. The fact that there are still discs observed in the vicinity of the star suggest that either the discs were initially massive ($\gtrsim 1 \mathrm{\ M}_{\odot}$), or that $\theta^{1}C$ Ori is younger than \SI{0.1}{Myr}. Given that such massive discs are gravitationally unstable and that the stellar age distribution in the ONC is $\sim \SI{1}{Myr}$ \citep[e.g.][]{dario2010, dario2012}, there must be other factors at play for these discs to have survived. \citet{storzer1999} argue that these discs are observed at a special moment in time: while coincidentally passing close to the center of the ONC, but they have spent the majority of their lives in regions of lower background radiation fields. \citet{scally2001} simulate the inner area of the ONC and propose that the type of orbits necessary for the hypothesis of \citet{storzer1999} to be feasible are not dynamically possible. However, a short-lived trajectory through high stellar density regions might not be the only way to explain the presence of massive discs in the center of the ONC. \citet{winter2019a} propose a combination of mechanisms, among which an extended period of star formation, that enable the presence of massive discs in the vicinity of massive stars, even at relatively late times. Our results also show that the high mass component of a disc population typically conists of young discs. 

Recently, several studies \citep{winter2020b, kruijssen2020a,
  chevance2021a, longmore2021} reported correlations between the
architecture of exoplanetary systems and the phase space densities in
the vicinity of their host stars. These results imply that external
influences can shape exoplanetary systems. The authors propose two
origins of these differences in phase space density: traces of the
star's phase space density within its birth region, or large-scale
galactic perturbations. These imply that exoplanetary systems are
shaped by processes in the birth cluster and in the galaxy,
respectively. Subsequent work by \citet{kruijssen2021} showed that the
phase space overdensities coincide with known structures driven by
large-scale galactic perturbations. This favours the galactic
perturbation scenario. The existence of influences on planet formation
due to cluster processes are not ruled out, but the galactic
population of field stars is sufficiently well-mixed to erase
signatures of the structure of the star-forming region. Also,
\citet{kruijssen2012} proposes that bound clusters arise from star
formation in regions of high ISM and stellar density, which implies
that the unbound population (which forms the field population)
consists of stars formed in low-density regions. The impact of a
high-density birth environment on planet formation can perhaps be
uncovered in the contrast between the exoplanetary systems of cluster
stars and field stars.

\section{Summary and conclusions}

We perform simulations of molecular cloud collapse and star formation. Our calculations begin with the collapse of a $10^4 \mathrm{\ M}_{\odot}$ molecular cloud with a 3\,pc radius. The star formation process ends when the star formation efficiency reaches 30\%. The excess gas is then instantaneously removed. From this point, we continue the simulations using a combination of $N$-body dynamics, dynamical truncation, viscous disc evolution, and stellar evolution. After formation stars with a mass $\mathrm{M}_* \leq 1.9 \mathrm{M}_{\odot}$ receive circumstellar discs. Stars more massive than $\mathrm{M}_* > 1.9 \mathrm{M}_{\odot}$ are emitting UV radiation, which affects the discs in their vicinity. We also take the internal photoevaporation of the host star into account, as well as the dynamical truncations and viscous evolution. We also implemented a model for photoevaporation of dust. We run the simulations for \SI{2}{Myr} after the last star has formed. We conclude that:

\begin{enumerate}[leftmargin=*]
\item[1.] A prolonged period of star formation, lasting from 2 to
  \SI{5}{Myr}, allows for relatively massive ($\mathrm{M}_\mathrm{gas}
  \sim 100 \mathrm{\ M}_{Jup}$, $\mathrm{M}_\mathrm{dust} \sim 100
  \mathrm{\ M}_{\oplus}$) discs to survive the high intensity UV
  radiation of nearby cluster members, and therewith provide a
  solution to the proplyd lifetime problem.  An extended periods of
  star formation then mediates the survival of massive discs in
  regions with a strong radiation field.
  
\item[2.] The discs that make it to the end of our simulations are
  preferentially the ones that were born later (after $\sim
  \SI{3}{Myr}$) in the star formation process.
  
\item[3.] While photoevaporation removes gas from the discs, dust
  quickly becomes resistant to photoevaporation. This might allow
  discs to keep enough mass to form rocky planets, even when depleted
  of gas.
  

\end{enumerate}

Photoevaporation due to the radiation of nearby massive stars is an
essential process that drives the dissolution of circumstellar discs.
The structure of the star cluster is then important in determining the
effectiveness of this process. Intracluster gas can shield discs from
this process, in which case stellar dynamical processes become
relatively more important.  Another important aspect of cluster
formation is the local star-formation history. Massive stars born late
may be unable to evaporate circumstellar discs effectively, because
the cluster, by that time, has expanded, ad the discs may already have
turned into planetary systems.  On the other hand, a late-formed disc
is protected by the larger distance to massive stars of the older
cluster.  Such lower density may be caused by the expulsion of the
primordial gas from the parent cluster.

These relatively fundamental processes in the star-forming environment give rise to a wide variety of disc masses and sizes that depend on the dynamical history of the parent star. Two identical stars with identical discs born in the same cluster but on a different orbit or born at a different time can then develop completely different disc parameters. The resulting planetary system will subsequently also be very different.
In our simulations we observe ranges in disc masses and sizes extending over more than three orders of magnitude, just from the time and orbit in which the disc-hosting star was born.

\section*{Acknowledgements}

It is a pleasure to thank the referee for detailed comments.  This project was supported by NOVA,
and we used the National Dutch Supercomputer Cartesius, and Little
Green Machine-II.

\section*{Data availability}
{\bf The code underlying the simulations in this article is available at \url{https://doi.org/10.5281/zenodo.6579534}.}

\section*{Software}
The present works makes use of the following software: AMUSE
\citep{portegieszwart2013, pelupessy2013}, Fi \citep{pelupessy2004},
ph4, Bridge \citep{fujii2007, portegieszwart2020}, SeBa
\citep{1996A&A...309..179P,2020A&A...640A..16T}, VADER
\citep{krumholz2015}, \texttt{numpy} \citep{vanderwalt2011a},
\texttt{scipy} \citep{virtanen2019}, \texttt{matplotlib}
\citep{hunter2007a}, and \texttt{makecite} \citep{price-whelan2018}.

\section*{Energy consumption}
The series of simulations presented in this work took $\sim$20 days to
run on 30 cores each. Although we present the results of 6 runs,
triple that number was eventually run (including code development,
testruns and referee requests for additional simulations). The total
amount of computer time used then tops 259.200 hours. Considering 12 Wh
for a CPU, this results in $\sim$3000 kWh. Using the conversion factor
0.283 kWh/kg \citep{portegieszwart2020a}, would results in
$\sim$10.000\,kg of CO$_2$. We used two computers for the calculations,
the Dutch National supercomputer Cartesius, and our local 192-core
workstation. Both machines are powered with renewable resources; the
former using Norwegian hydroelectric power (using certificates) and
the latter by Dutch wind-mill energy.




\bibliographystyle{mnras}
\bibliography{fconcha} 

\begin{thebibliography}{}
\makeatletter
\relax
\def\mn@urlcharsother{\let\do\@makeother \do\$\do\&\do\#\do\^\do\_\do\%\do\~}
\def\mn@doi{\begingroup\mn@urlcharsother \@ifnextchar [ {\mn@doi@}
  {\mn@doi@[]}}
\def\mn@doi@[#1]#2{\def\@tempa{#1}\ifx\@tempa\@empty \href
  {http://dx.doi.org/#2} {doi:#2}\else \href {http://dx.doi.org/#2} {#1}\fi
  \endgroup}
\def\mn@eprint#1#2{\mn@eprint@#1:#2::\@nil}
\def\mn@eprint@arXiv#1{\href {http://arxiv.org/abs/#1} {{\tt arXiv:#1}}}
\def\mn@eprint@dblp#1{\href {http://dblp.uni-trier.de/rec/bibtex/#1.xml}
  {dblp:#1}}
\def\mn@eprint@#1:#2:#3:#4\@nil{\def\@tempa {#1}\def\@tempb {#2}\def\@tempc
  {#3}\ifx \@tempc \@empty \let \@tempc \@tempb \let \@tempb \@tempa \fi \ifx
  \@tempb \@empty \def\@tempb {arXiv}\fi \@ifundefined
  {mn@eprint@\@tempb}{\@tempb:\@tempc}{\expandafter \expandafter \csname
  mn@eprint@\@tempb\endcsname \expandafter{\@tempc}}}

\bibitem[\protect\citeauthoryear{Adams}{Adams}{2010}]{adams2010}
Adams F.~C.,  2010, \mn@doi [\araa] {10.1146/annurev-astro-081309-130830}, 48,
  47

\bibitem[\protect\citeauthoryear{Adams, Hollenbach, Laughlin  \& Gorti}{Adams
  et~al.}{2004}]{adams2004}
Adams F.~C.,  Hollenbach D.,  Laughlin G.,   Gorti U.,  2004, \mn@doi [\apj]
  {10.1086/421989}, 611, 360

\bibitem[\protect\citeauthoryear{Adams, Proszkow, Fatuzzo  \& Myers}{Adams
  et~al.}{2006}]{adams2006}
Adams F.~C.,  Proszkow E.~M.,  Fatuzzo M.,   Myers P.~C.,  2006, \mn@doi [\apj]
  {10.1086/500393}, 641, 504

\bibitem[\protect\citeauthoryear{Anderson, Adams  \& Calvet}{Anderson
  et~al.}{2013}]{anderson2013}
Anderson K.~R.,  Adams F.~C.,   Calvet N.,  2013, \mn@doi [\apj]
  {10.1088/0004-637X/774/1/9}, 774, 9

\bibitem[\protect\citeauthoryear{Ansdell et~al.,}{Ansdell
  et~al.}{2016}]{ansdell2016}
Ansdell M.,  et~al., 2016, \mn@doi [\apj] {10.3847/0004-637X/828/1/46}, 828, 46

\bibitem[\protect\citeauthoryear{Ansdell, Williams, Manara, Miotello, Facchini,
  van~der Marel, Testi  \& van Dishoeck}{Ansdell et~al.}{2017}]{ansdell2017}
Ansdell M.,  Williams J.~P.,  Manara C.~F.,  Miotello A.,  Facchini S.,
  van~der Marel N.,  Testi L.,   van Dishoeck E.~F.,  2017, \mn@doi [\aj]
  {10.3847/1538-3881/aa69c0}, 153, 240

\bibitem[\protect\citeauthoryear{Ansdell et~al.,}{Ansdell
  et~al.}{2020}]{ansdell2020}
Ansdell M.,  et~al., 2020, \mn@doi [\aj] {10.3847/1538-3881/abb9af}, 160, 248

\bibitem[\protect\citeauthoryear{Armitage}{Armitage}{2000}]{armitage2000}
Armitage P.~J.,  2000, \aap, 362, 968

\bibitem[\protect\citeauthoryear{Balog, Muzerolle, Rieke, Su, Young  \&
  Megeath}{Balog et~al.}{2007}]{balog2007}
Balog Z.,  Muzerolle J.,  Rieke G.~H.,  Su K. Y.~L.,  Young E.~T.,   Megeath
  S.~T.,  2007, \mn@doi [\apj] {10.1086/513311}, 660, 1532

\bibitem[\protect\citeauthoryear{Banerjee \& Kroupa}{Banerjee \&
  Kroupa}{2015}]{banerjee2015}
Banerjee S.,  Kroupa P.,  2015, \mn@doi [Monthly Notices of the Royal
  Astronomical Society] {10.1093/mnras/stu2445}, 447, 728

\bibitem[\protect\citeauthoryear{Bate, Bonnell  \& Bromm}{Bate
  et~al.}{2003}]{bate2003}
Bate M.~R.,  Bonnell I.~A.,   Bromm V.,  2003, \mn@doi [\mnras]
  {10.1046/j.1365-8711.2003.06210.x}, 339, 577

\bibitem[\protect\citeauthoryear{Birnstiel, Dullemond  \& Brauer}{Birnstiel
  et~al.}{2010}]{birnstiel2010}
Birnstiel T.,  Dullemond C.~P.,   Brauer F.,  2010, \mn@doi [\aap]
  {10.1051/0004-6361/200913731}, 513, A79

\bibitem[\protect\citeauthoryear{Bohlin, Savage  \& Drake}{Bohlin
  et~al.}{1978}]{bohlin1978}
Bohlin R.~C.,  Savage B.~D.,   Drake J.~F.,  1978, \mn@doi [\apj]
  {10.1086/156357}, 224, 132

\bibitem[\protect\citeauthoryear{Bontemps et~al.,}{Bontemps
  et~al.}{2001}]{bontemps2001}
Bontemps S.,  et~al., 2001, \mn@doi [\aap] {10.1051/0004-6361:20010474}, 372,
  173

\bibitem[\protect\citeauthoryear{Breslau, Steinhausen, Vincke  \&
  Pfalzner}{Breslau et~al.}{2014}]{breslau2014}
Breslau A.,  Steinhausen M.,  Vincke K.,   Pfalzner S.,  2014, \mn@doi [\aap]
  {10.1051/0004-6361/201323043}, 565, A130

\bibitem[\protect\citeauthoryear{Cabrit, Pety, Pesenti  \& Dougados}{Cabrit
  et~al.}{2006}]{cabrit2006}
Cabrit S.,  Pety J.,  Pesenti N.,   Dougados C.,  2006, \mn@doi [\aap]
  {10.1051/0004-6361:20054047}, 452, 897

\bibitem[\protect\citeauthoryear{Carpenter, Mamajek, Hillenbrand  \&
  Meyer}{Carpenter et~al.}{2006}]{carpenter2006}
Carpenter J.~M.,  Mamajek E.~E.,  Hillenbrand L.~A.,   Meyer M.~R.,  2006,
  \mn@doi [\apj] {10.1086/509121}, 651, L49

\bibitem[\protect\citeauthoryear{Cartwright \& Whitworth}{Cartwright \&
  Whitworth}{2004}]{cartwright2004}
Cartwright A.,  Whitworth A.~P.,  2004, \mn@doi [\mnras]
  {10.1111/j.1365-2966.2004.07360.x}, 348, 589

\bibitem[\protect\citeauthoryear{Casertano \& Hut}{Casertano \&
  Hut}{1985}]{casertano1985}
Casertano S.,  Hut P.,  1985, \mn@doi [\apj] {10.1086/163589}, 298, 80

\bibitem[\protect\citeauthoryear{Chevance et~al.,}{Chevance
  et~al.}{2020}]{chevance2020a}
Chevance M.,  et~al., 2020, \mn@doi [\textbackslash mnras]
  {10.1093/mnras/stz3525}, 493, 2872

\bibitem[\protect\citeauthoryear{Chevance, Kruijssen  \& Longmore}{Chevance
  et~al.}{2021}]{chevance2021a}
Chevance M.,  Kruijssen J. M.~D.,   Longmore S.~N.,  2021, arXiv e-prints,
  2103, arXiv:2103.08604

\bibitem[\protect\citeauthoryear{Clarke}{Clarke}{2007}]{clarke2007}
Clarke C.~J.,  2007, \mn@doi [\mnras] {10.1111/j.1365-2966.2007.11547.x}, 376,
  1350

\bibitem[\protect\citeauthoryear{Clarke \& Pringle}{Clarke \&
  Pringle}{1991}]{clarke1991}
Clarke C.~J.,  Pringle J.~E.,  1991, \mn@doi [\mnras]
  {10.1093/mnras/249.4.588}, 249, 588

\bibitem[\protect\citeauthoryear{Clarke \& Pringle}{Clarke \&
  Pringle}{1993}]{clarke1993}
Clarke C.~J.,  Pringle J.~E.,  1993, \mn@doi [\mnras]
  {10.1093/mnras/261.1.190}, 261, 190

\bibitem[\protect\citeauthoryear{{Cleeves}, {{\"O}berg}, {Wilner}, {Huang},
  {Loomis}, {Andrews}  \& {Czekala}}{{Cleeves} et~al.}{2016}]{cleeves2016}
{Cleeves} L.~I.,  {{\"O}berg} K.~I.,  {Wilner} D.~J.,  {Huang} J.,  {Loomis}
  R.~A.,  {Andrews} S.~M.,   {Czekala} I.,  2016, \mn@doi [\apj]
  {10.3847/0004-637X/832/2/110}, \href
  {https://ui.adsabs.harvard.edu/abs/2016ApJ...832..110C} {832, 110}

\bibitem[\protect\citeauthoryear{Concha-Ram\'irez, Vaher  \&
  Portegies~Zwart}{Concha-Ram\'irez et~al.}{2019a}]{concha-ramirez2019}
Concha-Ram\'irez F.,  Vaher E.,   Portegies~Zwart S.,  2019a, \mn@doi [\mnras]
  {10.1093/mnras/sty2721}, 482, 732

\bibitem[\protect\citeauthoryear{Concha-Ram\'irez, Wilhelm, Portegies~Zwart  \&
  Haworth}{Concha-Ram\'irez et~al.}{2019b}]{concha-ramirez2019a}
Concha-Ram\'irez F.,  Wilhelm M. J.~C.,  Portegies~Zwart S.,   Haworth T.~J.,
  2019b, \mn@doi [\mnras] {10.1093/mnras/stz2973}, 490, 5678

\bibitem[\protect\citeauthoryear{Concha-Ram\'irez, Wilhelm, Zwart, van Terwisga
   \& Hacar}{Concha-Ram\'irez et~al.}{2021}]{concha-ramirez2021}
Concha-Ram\'irez F.,  Wilhelm M. J.~C.,  Zwart S.~P.,  van Terwisga S.~E.,
  Hacar A.,  2021, \mn@doi [\mnras] {10.1093/mnras/staa3669}, 501, 1782

\bibitem[\protect\citeauthoryear{Cuello et~al.,}{Cuello
  et~al.}{2018}]{cuello2018}
Cuello N.,  et~al., 2018, Proceedings of the Annual meeting of the French
  Society of Astronomy and Astrophysics, pp 95--101

\bibitem[\protect\citeauthoryear{Da~Rio, Robberto, Soderblom, Panagia,
  Hillenbrand, Palla  \& Stassun}{Da~Rio et~al.}{2010}]{dario2010}
Da~Rio N.,  Robberto M.,  Soderblom D.~R.,  Panagia N.,  Hillenbrand L.~A.,
  Palla F.,   Stassun K.~G.,  2010, \mn@doi [\apj]
  {10.1088/0004-637X/722/2/1092}, 722, 1092

\bibitem[\protect\citeauthoryear{Da~Rio, Robberto, Hillenbrand, Henning  \&
  Stassun}{Da~Rio et~al.}{2012}]{dario2012}
Da~Rio N.,  Robberto M.,  Hillenbrand L.~A.,  Henning T.,   Stassun K.~G.,
  2012, \mn@doi [\apj] {10.1088/0004-637X/748/1/14}, 748, 14

\bibitem[\protect\citeauthoryear{Eisner \& Carpenter}{Eisner \&
  Carpenter}{2006}]{eisner2006}
Eisner J.~A.,  Carpenter J.~M.,  2006, \mn@doi [\apj] {10.1086/500637}, 641,
  1162

\bibitem[\protect\citeauthoryear{Eisner, Plambeck, Carpenter, Corder, Qi  \&
  Wilner}{Eisner et~al.}{2008}]{eisner2008}
Eisner J.~A.,  Plambeck R.~L.,  Carpenter J.~M.,  Corder S.~A.,  Qi C.,
  Wilner D.,  2008, \mn@doi [\apj] {10.1086/588524}, 683, 304

\bibitem[\protect\citeauthoryear{Eisner et~al.,}{Eisner
  et~al.}{2018}]{eisner2018}
Eisner J.~A.,  et~al., 2018, \mn@doi [\apj] {10.3847/1538-4357/aac3e2}, 860, 77

\bibitem[\protect\citeauthoryear{Elmegreen}{Elmegreen}{2008}]{elmegreen2008}
Elmegreen B.~G.,  2008, \mn@doi [The Astrophysical Journal] {10.1086/523791},
  672, 1006

\bibitem[\protect\citeauthoryear{Elmegreen, Efremov, Pudritz  \&
  Zinnecker}{Elmegreen et~al.}{2000}]{elmegreen2000}
Elmegreen B.~G.,  Efremov Y.,  Pudritz R.~E.,   Zinnecker H.,  2000, Protostars
  and Planets IV, p.~179

\bibitem[\protect\citeauthoryear{Facchini, Clarke  \& Bisbas}{Facchini
  et~al.}{2016}]{facchini2016}
Facchini S.,  Clarke C.~J.,   Bisbas T.~G.,  2016, \mn@doi [\mnras]
  {10.1093/mnras/stw240}, 457, 3593

\bibitem[\protect\citeauthoryear{Falgarone \& Phillips}{Falgarone \&
  Phillips}{1991}]{falgarone1991a}
Falgarone E.,  Phillips T.~G.,  1991, Proceedings of the International
  Astronomical Union, 147, 119

\bibitem[\protect\citeauthoryear{Falgarone, Phillips  \& Walker}{Falgarone
  et~al.}{1991}]{falgarone1991}
Falgarone E.,  Phillips T.~G.,   Walker C.~K.,  1991, \mn@doi [\apj]
  {10.1086/170419}, 378, 186

\bibitem[\protect\citeauthoryear{Fang et~al.,}{Fang et~al.}{2012}]{fang2012}
Fang M.,  et~al., 2012, \mn@doi [\aap] {10.1051/0004-6361/201015914}, 539, A119

\bibitem[\protect\citeauthoryear{Fatuzzo \& Adams}{Fatuzzo \&
  Adams}{2008}]{fatuzzo2008}
Fatuzzo M.,  Adams F.~C.,  2008, \mn@doi [\apj] {10.1086/527469}, 675, 1361

\bibitem[\protect\citeauthoryear{Flaccomio, Micela  \& Sciortino}{Flaccomio
  et~al.}{2012}]{flaccomio2012}
Flaccomio E.,  Micela G.,   Sciortino S.,  2012, \mn@doi [\aap]
  {10.1051/0004-6361/201219362}, 548, A85

\bibitem[\protect\citeauthoryear{Font, McCarthy, Johnstone  \& Ballantyne}{Font
  et~al.}{2004}]{font2004}
Font A.~S.,  McCarthy I.~G.,  Johnstone D.,   Ballantyne D.~R.,  2004, \mn@doi
  [\apj] {10.1086/383518}, 607, 890

\bibitem[\protect\citeauthoryear{{Fujii} \& {Portegies Zwart}}{{Fujii} \&
  {Portegies Zwart}}{2016}]{2016ApJ...817....4F}
{Fujii} M.~S.,  {Portegies Zwart} S.,  2016, \mn@doi [\apj]
  {10.3847/0004-637X/817/1/4}, \href
  {http://adsabs.harvard.edu/abs/2016ApJ...817....4F} {817, 4}

\bibitem[\protect\citeauthoryear{Fujii, Iwasawa, Funato  \& Makino}{Fujii
  et~al.}{2007}]{fujii2007}
Fujii M.,  Iwasawa M.,  Funato Y.,   Makino J.,  2007, \mn@doi [Publications of
  the Astronomical Society of Japan] {10.1093/pasj/59.6.1095}, 59, 1095

\bibitem[\protect\citeauthoryear{Galli, Bouy, Olivares, Miret-Roig, Sarro,
  Barrado, Berihuete  \& Brandner}{Galli et~al.}{2020}]{galli2020}
Galli P. A.~B.,  Bouy H.,  Olivares J.,  Miret-Roig N.,  Sarro L.~M.,  Barrado
  D.,  Berihuete A.,   Brandner W.,  2020, \mn@doi [\aap]
  {10.1051/0004-6361/201936708}, 634, A98

\bibitem[\protect\citeauthoryear{Goodwin \& Whitworth}{Goodwin \&
  Whitworth}{2004}]{goodwin2004}
Goodwin S.~P.,  Whitworth A.~P.,  2004, \mn@doi [\aap]
  {10.1051/0004-6361:20031529}, 413, 929

\bibitem[\protect\citeauthoryear{Gorti \& Hollenbach}{Gorti \&
  Hollenbach}{2009}]{gorti2009a}
Gorti U.,  Hollenbach D.,  2009, \mn@doi [\apj] {10.1088/0004-637X/690/2/1539},
  690, 1539

\bibitem[\protect\citeauthoryear{Gorti, Dullemond  \& Hollenbach}{Gorti
  et~al.}{2009}]{gorti2009}
Gorti U.,  Dullemond C.~P.,   Hollenbach D.,  2009, \mn@doi [\apj]
  {10.1088/0004-637X/705/2/1237}, 705, 1237–1251

\bibitem[\protect\citeauthoryear{Guarcello et~al.,}{Guarcello
  et~al.}{2016}]{guarcello2016}
Guarcello M.~G.,  et~al., 2016, arXiv:1605.01773 [astro-ph]

\bibitem[\protect\citeauthoryear{Habing}{Habing}{1968}]{habing1968}
Habing H.~J.,  1968, Bulletin of the Astronomical Institutes of the
  Netherlands, 19, 421

\bibitem[\protect\citeauthoryear{Hacar, Tafalla, Forbrich, Alves, Meingast,
  Grossschedl  \& Teixeira}{Hacar et~al.}{2018}]{hacar2018}
Hacar A.,  Tafalla M.,  Forbrich J.,  Alves J.,  Meingast S.,  Grossschedl J.,
   Teixeira P.~S.,  2018, \mn@doi [\aap] {10.1051/0004-6361/201731894}, 610,
  A77

\bibitem[\protect\citeauthoryear{Hansen, Klein, McKee  \& Fisher}{Hansen
  et~al.}{2012}]{hansen2012}
Hansen C.~E.,  Klein R.~I.,  McKee C.~F.,   Fisher R.~T.,  2012, \mn@doi [\apj]
  {10.1088/0004-637X/747/1/22}, 747, 22

\bibitem[\protect\citeauthoryear{Hartmann}{Hartmann}{2002}]{hartmann2002}
Hartmann L.,  2002, \mn@doi [\apj] {10.1086/342657}, 578, 914

\bibitem[\protect\citeauthoryear{Haworth \& Clarke}{Haworth \&
  Clarke}{2019}]{haworth2019}
Haworth T.~J.,  Clarke C.~J.,  2019, \mn@doi [\mnras] {10.1093/mnras/stz706},
  485, 3895

\bibitem[\protect\citeauthoryear{Haworth, Boubert, Facchini, Bisbas  \&
  Clarke}{Haworth et~al.}{2016}]{haworth2016}
Haworth T.~J.,  Boubert D.,  Facchini S.,  Bisbas T.~G.,   Clarke C.~J.,  2016,
  \mn@doi [\mnras] {10.1093/mnras/stw2280}, 463, 3616

\bibitem[\protect\citeauthoryear{Haworth, Facchini, Clarke  \& Cleeves}{Haworth
  et~al.}{2017}]{haworth2017}
Haworth T.~J.,  Facchini S.,  Clarke C.~J.,   Cleeves L.~I.,  2017, \mn@doi
  [\mnras] {10.1093/mnrasl/slx037}, 468, L108

\bibitem[\protect\citeauthoryear{Haworth, Facchini, Clarke  \& Mohanty}{Haworth
  et~al.}{2018a}]{haworth2018a}
Haworth T.~J.,  Facchini S.,  Clarke C.~J.,   Mohanty S.,  2018a, \mn@doi
  [\mnras] {10.1093/mnras/sty168}, 475, 5460

\bibitem[\protect\citeauthoryear{Haworth, Clarke, Rahman, Winter  \&
  Facchini}{Haworth et~al.}{2018b}]{haworth2018}
Haworth T.~J.,  Clarke C.~J.,  Rahman W.,  Winter A.~J.,   Facchini S.,  2018b,
  \mn@doi [\mnras] {10.1093/mnras/sty2323}, 481, 452

\bibitem[\protect\citeauthoryear{Hillenbrand \& Hartmann}{Hillenbrand \&
  Hartmann}{1998}]{hillenbrand1998}
Hillenbrand L.~A.,  Hartmann L.~W.,  1998, \mn@doi [\apj] {10.1086/305076},
  492, 540

\bibitem[\protect\citeauthoryear{Hollenbach, Yorke  \& Johnstone}{Hollenbach
  et~al.}{2000}]{hollenbach2000}
Hollenbach D.~J.,  Yorke H.~W.,   Johnstone D.,  2000, Protostars and Planets
  IV, p.~401

\bibitem[\protect\citeauthoryear{Hunter}{Hunter}{2007}]{hunter2007a}
Hunter J.~D.,  2007, \mn@doi [Computing in Science & Engineering]
  {10.1109/MCSE.2007.55}, 9, 90

\bibitem[\protect\citeauthoryear{J\'ilkov\'a, Portegies~Zwart, Pijloo  \&
  Hammer}{J\'ilkov\'a et~al.}{2015}]{jilkova2015}
J\'ilkov\'a L.,  Portegies~Zwart S.,  Pijloo T.,   Hammer M.,  2015, \mn@doi
  [\mnras] {10.1093/mnras/stv1803}, 453, 3157

\bibitem[\protect\citeauthoryear{J\'ilkov\'a, Hamers, Hammer  \&
  Portegies~Zwart}{J\'ilkov\'a et~al.}{2016}]{jilkova2016}
J\'ilkov\'a L.,  Hamers A.~S.,  Hammer M.,   Portegies~Zwart S.,  2016, \mn@doi
  [\mnras] {10.1093/mnras/stw264}, 457, 4218

\bibitem[\protect\citeauthoryear{Johnstone, Hollenbach  \& Bally}{Johnstone
  et~al.}{1998}]{johnstone1998}
Johnstone D.,  Hollenbach D.,   Bally J.,  1998, \mn@doi [\apj]
  {10.1086/305658}, 499, 758

\bibitem[\protect\citeauthoryear{Kim, Clarke, Fang  \& Facchini}{Kim
  et~al.}{2016}]{kim2016}
Kim J.~S.,  Clarke C.~J.,  Fang M.,   Facchini S.,  2016, \mn@doi [\apj]
  {10.3847/2041-8205/826/1/L15}, 826, L15

\bibitem[\protect\citeauthoryear{Kraus \& Hillenbrand}{Kraus \&
  Hillenbrand}{2008}]{kraus2008}
Kraus A.~L.,  Hillenbrand L.~A.,  2008, \mn@doi [\apjl] {10.1086/593012}, 686,
  L111

\bibitem[\protect\citeauthoryear{Kroupa}{Kroupa}{2001}]{kroupa2001}
Kroupa P.,  2001, \mn@doi [\mnras] {10.1046/j.1365-8711.2001.04022.x}, 322, 231

\bibitem[\protect\citeauthoryear{Kruijssen}{Kruijssen}{2012}]{kruijssen2012a}
Kruijssen J. M.~D.,  2012, \mn@doi [Monthly Notices of the Royal Astronomical
  Society] {10.1111/j.1365-2966.2012.21923.x}, 426, 3008

\bibitem[\protect\citeauthoryear{Kruijssen, Maschberger, Moeckel, Clarke,
  Bastian  \& Bonnell}{Kruijssen et~al.}{2012}]{kruijssen2012}
Kruijssen J. M.~D.,  Maschberger T.,  Moeckel N.,  Clarke C.~J.,  Bastian N.,
  Bonnell I.~A.,  2012, \mn@doi [\mnras] {10.1111/j.1365-2966.2011.19748.x},
  419, 841

\bibitem[\protect\citeauthoryear{Kruijssen, Longmore  \& Chevance}{Kruijssen
  et~al.}{2020}]{kruijssen2020a}
Kruijssen J. M.~D.,  Longmore S.~N.,   Chevance M.,  2020, \mn@doi [The
  Astrophysical Journal Letters] {10.3847/2041-8213/abccc3}, 905, L18

\bibitem[\protect\citeauthoryear{{Kruijssen}, {Longmore}, {Chevance},
  {Laporte}, {Motylinski}, {Keller}  \& {Henshaw}}{{Kruijssen}
  et~al.}{2021}]{kruijssen2021}
{Kruijssen} J.~M.~D.,  {Longmore} S.~N.,  {Chevance} M.,  {Laporte} C. F.~P.,
  {Motylinski} M.,  {Keller} B.~W.,   {Henshaw} J.~D.,  2021, arXiv e-prints,
  \href {https://ui.adsabs.harvard.edu/abs/2021arXiv210906182K} {p.
  arXiv:2109.06182}

\bibitem[\protect\citeauthoryear{{Krumholz}}{{Krumholz}}{2014}]{krumholz2014}
{Krumholz} M.~R.,  2014, \mn@doi [\physrep] {10.1016/j.physrep.2014.02.001},
  \href {https://ui.adsabs.harvard.edu/abs/2014PhR...539...49K} {539, 49}

\bibitem[\protect\citeauthoryear{Krumholz \& Forbes}{Krumholz \&
  Forbes}{2015}]{krumholz2015}
Krumholz M.~R.,  Forbes J.~C.,  2015, \mn@doi [Astronomy and Computing]
  {10.1016/j.ascom.2015.02.005}, 11, 1

\bibitem[\protect\citeauthoryear{Krumholz, McKee  \& Bland-Hawthorn}{Krumholz
  et~al.}{2019}]{krumholz2019}
Krumholz M.~R.,  McKee C.~F.,   Bland-Hawthorn J.,  2019, \mn@doi [\araa]
  {10.1146/annurev-astro-091918-104430}, 57, 227

\bibitem[\protect\citeauthoryear{Kuhn, Hillenbrand, Sills, Feigelson  \&
  Getman}{Kuhn et~al.}{2019}]{kuhn2019}
Kuhn M.~A.,  Hillenbrand L.~A.,  Sills A.,  Feigelson E.~D.,   Getman K.~V.,
  2019, \mn@doi [\apj] {10.3847/1538-4357/aaef8c}, 870, 32

\bibitem[\protect\citeauthoryear{Lada \& Lada}{Lada \& Lada}{2003}]{lada2003}
Lada C.~J.,  Lada E.~A.,  2003, \mn@doi [\araa]
  {10.1146/annurev.astro.41.011802.094844}, 41, 57

\bibitem[\protect\citeauthoryear{Larson}{Larson}{1981}]{larson1981}
Larson R.~B.,  1981, \mn@doi [Monthly Notices of the Royal Astronomical
  Society] {10.1093/mnras/194.4.809}, 194, 809

\bibitem[\protect\citeauthoryear{Larson}{Larson}{1995}]{larson1995}
Larson R.~B.,  1995, \mn@doi [\mnras] {10.1093/mnras/272.1.213}, 272, 213

\bibitem[\protect\citeauthoryear{Longmore, Chevance  \& Kruijssen}{Longmore
  et~al.}{2021}]{longmore2021}
Longmore S.~N.,  Chevance M.,   Kruijssen J. M.~D.,  2021, arXiv e-prints,
  2103, arXiv:2103.01974

\bibitem[\protect\citeauthoryear{Lopez \& Rice}{Lopez \&
  Rice}{2018}]{lopez2018}
Lopez E.~D.,  Rice K.,  2018, \mn@doi [Monthly Notices of the Royal
  Astronomical Society] {10.1093/mnras/sty1707}, 479, 5303

\bibitem[\protect\citeauthoryear{Luhman}{Luhman}{2007}]{luhman2007}
Luhman K.~L.,  2007, \mn@doi [\apjs] {10.1086/520114}, 173, 104

\bibitem[\protect\citeauthoryear{Luhman}{Luhman}{2018}]{luhman2018}
Luhman K.~L.,  2018, \mn@doi [\aj] {10.3847/1538-3881/aae831}, 156, 271

\bibitem[\protect\citeauthoryear{Luhman \& Esplin}{Luhman \&
  Esplin}{2020}]{luhman2020}
Luhman K.~L.,  Esplin T.~L.,  2020, \mn@doi [\aj] {10.3847/1538-3881/ab9599},
  160, 44

\bibitem[\protect\citeauthoryear{Luhman \& Mamajek}{Luhman \&
  Mamajek}{2012}]{luhman2012}
Luhman K.~L.,  Mamajek E.~E.,  2012, \mn@doi [\apj]
  {10.1088/0004-637X/758/1/31}, 758, 31

\bibitem[\protect\citeauthoryear{Luhman, Esplin  \& Loutrel}{Luhman
  et~al.}{2016}]{luhman2016}
Luhman K.~L.,  Esplin T.~L.,   Loutrel N.~P.,  2016, \mn@doi [\apj]
  {10.3847/0004-637X/827/1/52}, 827, 52

\bibitem[\protect\citeauthoryear{Lynden-Bell \& Pringle}{Lynden-Bell \&
  Pringle}{1974}]{lynden-bell1974}
Lynden-Bell D.,  Pringle J.~E.,  1974, \mn@doi [\mnras]
  {10.1093/mnras/168.3.603}, 168, 603

\bibitem[\protect\citeauthoryear{Mac~Low \& Klessen}{Mac~Low \&
  Klessen}{2004}]{maclow2004}
Mac~Low M.-M.,  Klessen R.~S.,  2004, \mn@doi [Reviews of Modern Physics]
  {10.1103/RevModPhys.76.125}, 76, 125

\bibitem[\protect\citeauthoryear{Manara et~al.,}{Manara
  et~al.}{2020}]{manara2020}
Manara C.~F.,  et~al., 2020, \mn@doi [\aap] {10.1051/0004-6361/202037949}, 639,
  A58

\bibitem[\protect\citeauthoryear{Mann \& Williams}{Mann \&
  Williams}{2010}]{mann2010}
Mann R.~K.,  Williams J.~P.,  2010, \mn@doi [\apj]
  {10.1088/0004-637X/725/1/430}, 725, 430

\bibitem[\protect\citeauthoryear{Mann et~al.,}{Mann et~al.}{2014}]{mann2014}
Mann R.~K.,  et~al., 2014, \mn@doi [\apj] {10.1088/0004-637X/784/1/82}, 784, 82

\bibitem[\protect\citeauthoryear{Matzner \& McKee}{Matzner \&
  McKee}{2000}]{matzner2000}
Matzner C.~D.,  McKee C.~F.,  2000, \mn@doi [\apj] {10.1086/317785}, 545, 364

\bibitem[\protect\citeauthoryear{Neuh\"auser \& Forbrich}{Neuh\"auser \&
  Forbrich}{2008}]{neuhauser2008}
Neuh\"auser R.,  Forbrich J.,  2008, Handbook of Star Forming Regions, Volume
  II, p.~735

\bibitem[\protect\citeauthoryear{Nicholson, Parker, Church, Davies, Fearon  \&
  Walton}{Nicholson et~al.}{2019}]{nicholson2019}
Nicholson R.~B.,  Parker R.~J.,  Church R.~P.,  Davies M.~B.,  Fearon N.~M.,
  Walton S. R.~J.,  2019, \mn@doi [\mnras] {10.1093/mnras/stz606}, 485, 4893

\bibitem[\protect\citeauthoryear{O'dell}{O'dell}{1998}]{odell1998}
O'dell C.~R.,  1998, \mn@doi [\aj] {10.1086/300178}, 115, 263

\bibitem[\protect\citeauthoryear{O'dell \& Wen}{O'dell \&
  Wen}{1994}]{odell1994}
O'dell C.~R.,  Wen Z.,  1994, \mn@doi [\mnras] {10.1086/174892}, 436, 194

\bibitem[\protect\citeauthoryear{Offner \& Arce}{Offner \&
  Arce}{2015}]{offner2015}
Offner S. S.~R.,  Arce H.~G.,  2015, \mn@doi [\apj]
  {10.1088/0004-637X/811/2/146}, 811, 146

\bibitem[\protect\citeauthoryear{Owen, Ercolano, Clarke  \& Alexander}{Owen
  et~al.}{2010}]{owen2010}
Owen J.~E.,  Ercolano B.,  Clarke C.~J.,   Alexander R.~D.,  2010, \mn@doi
  [\mnras] {10.1111/j.1365-2966.2009.15771.x}, 401, 1415

\bibitem[\protect\citeauthoryear{Owen, Clarke  \& Ercolano}{Owen
  et~al.}{2012}]{owen2012}
Owen J.~E.,  Clarke C.~J.,   Ercolano B.,  2012, \mn@doi [\mnras]
  {10.1111/j.1365-2966.2011.20337.x}, 422, 1880

\bibitem[\protect\citeauthoryear{Parker}{Parker}{2014}]{parker2014b}
Parker R.~J.,  2014, \mn@doi [\mnras] {10.1093/mnras/stu2054}, 445, 4037

\bibitem[\protect\citeauthoryear{Parker \& Alves~de Oliveira}{Parker \&
  Alves~de Oliveira}{2017}]{parker2017}
Parker R.~J.,  Alves~de Oliveira C.,  2017, \mn@doi [\mnras]
  {10.1093/mnras/stx739}, 468, 4340

\bibitem[\protect\citeauthoryear{Parker \& Meyer}{Parker \&
  Meyer}{2012}]{parker2012}
Parker R.~J.,  Meyer M.~R.,  2012, \mn@doi [\mnras]
  {10.1111/j.1365-2966.2012.21851.x}, 427, 637

\bibitem[\protect\citeauthoryear{Pascucci et~al.,}{Pascucci
  et~al.}{2016}]{pascucci2016}
Pascucci I.,  et~al., 2016, \mn@doi [\apj] {10.3847/0004-637X/831/2/125}, 831,
  125

\bibitem[\protect\citeauthoryear{Pelupessy, van~der Werf  \& Icke}{Pelupessy
  et~al.}{2004}]{pelupessy2004}
Pelupessy F.~I.,  van~der Werf P.~P.,   Icke V.,  2004, \mn@doi [\aap]
  {10.1051/0004-6361:20047071}, 422, 55

\bibitem[\protect\citeauthoryear{Pelupessy, van Elteren, de Vries, McMillan,
  Drost  \& Portegies~Zwart}{Pelupessy et~al.}{2013}]{pelupessy2013}
Pelupessy F.~I.,  van Elteren A.,  de Vries N.,  McMillan S. L.~W.,  Drost N.,
   Portegies~Zwart S.~F.,  2013, \mn@doi [\aap] {10.1051/0004-6361/201321252},
  557, A84

\bibitem[\protect\citeauthoryear{Petigura, Howard  \& Marcy}{Petigura
  et~al.}{2013}]{petigura2013}
Petigura E.~A.,  Howard A.~W.,   Marcy G.~W.,  2013, \mn@doi [Proceedings of
  the National Academy of Science] {10.1073/pnas.1319909110}, 110, 19273

\bibitem[\protect\citeauthoryear{Pfalzner}{Pfalzner}{2003}]{pfalzner2003}
Pfalzner S.,  2003, \mn@doi [\apj] {10.1086/375808}, 592, 986

\bibitem[\protect\citeauthoryear{Pfalzner, Vogel, Scharw\"achter  \&
  Olczak}{Pfalzner et~al.}{2005a}]{pfalzner2005a}
Pfalzner S.,  Vogel P.,  Scharw\"achter J.,   Olczak C.,  2005a, \mn@doi [\aap]
  {10.1051/0004-6361:20042467}, 437, 967

\bibitem[\protect\citeauthoryear{Pfalzner, Umbreit  \& Henning}{Pfalzner
  et~al.}{2005b}]{pfalzner2005}
Pfalzner S.,  Umbreit S.,   Henning T.,  2005b, \mn@doi [\apj]
  {10.1086/431350}, 629, 526

\bibitem[\protect\citeauthoryear{Pfalzner, Tackenberg  \& Steinhausen}{Pfalzner
  et~al.}{2008}]{pfalzner2008}
Pfalzner S.,  Tackenberg J.,   Steinhausen M.,  2008, \mn@doi [\aap]
  {10.1051/0004-6361:200810223}, 487, L45

\bibitem[\protect\citeauthoryear{Picogna, Ercolano, Owen  \& Weber}{Picogna
  et~al.}{2019}]{picogna2019}
Picogna G.,  Ercolano B.,  Owen J.~E.,   Weber M.~L.,  2019, \mn@doi [\mnras]
  {10.1093/mnras/stz1166}, 487, 691

\bibitem[\protect\citeauthoryear{Portegies~Zwart}{Portegies~Zwart}{2016}]{portegieszwart2016}
Portegies~Zwart S.~F.,  2016, \mn@doi [\mnras] {10.1093/mnras/stv2831}, 457,
  313

\bibitem[\protect\citeauthoryear{Portegies~Zwart}{Portegies~Zwart}{2020}]{portegieszwart2020a}
Portegies~Zwart S.,  2020, \mn@doi [Nature Astronomy]
  {10.1038/s41550-020-1208-y}, 4, 819

\bibitem[\protect\citeauthoryear{{Portegies Zwart} \& {Verbunt}}{{Portegies
  Zwart} \& {Verbunt}}{1996}]{1996A&A...309..179P}
{Portegies Zwart} S.~F.,  {Verbunt} F.,  1996, \aap, \href
  {http://adsabs.harvard.edu/cgi-bin/nph-bib_query?bibcode=1996A%26A...309..179P&amp;db_key=AST}
  {309, 179}

\bibitem[\protect\citeauthoryear{Portegies~Zwart, McMillan, van Elteren,
  Pelupessy  \& de Vries}{Portegies~Zwart et~al.}{2013}]{portegieszwart2013}
Portegies~Zwart S.,  McMillan S. L.~W.,  van Elteren E.,  Pelupessy I.,   de
  Vries N.,  2013, \mn@doi [Computer Physics Communications]
  {10.1016/j.cpc.2012.09.024}, 183, 456

\bibitem[\protect\citeauthoryear{Portegies~Zwart, Pelupessy,
  Mart\'inez-Barbosa, van Elteren  \& McMillan}{Portegies~Zwart
  et~al.}{2020}]{portegieszwart2020}
Portegies~Zwart S.,  Pelupessy I.,  Mart\'inez-Barbosa C.,  van Elteren A.,
  McMillan S.,  2020, \mn@doi [Communications in Nonlinear Science and
  Numerical Simulations] {10.1016/j.cnsns.2020.105240}, 85, 105240

\bibitem[\protect\citeauthoryear{Price-Whelan, Mechev  \&
  jumeroag}{Price-Whelan et~al.}{2018}]{price-whelan2018}
Price-Whelan A.,  Mechev A.,   jumeroag 2018, adrn/makecite: v0.2, Zenodo,
  \mn@doi{10.5281/zenodo.1343299}

\bibitem[\protect\citeauthoryear{Punzo, Capuzzo-Dolcetta  \&
  Portegies~Zwart}{Punzo et~al.}{2014}]{punzo2014}
Punzo D.,  Capuzzo-Dolcetta R.,   Portegies~Zwart S.,  2014, \mn@doi [\mnras]
  {10.1093/mnras/stu1650}, 444, 2808

\bibitem[\protect\citeauthoryear{Reche, Beust  \& Augereau}{Reche
  et~al.}{2009}]{reche2009}
Reche R.,  Beust H.,   Augereau J.-C.,  2009, \mn@doi [\aap]
  {10.1051/0004-6361:200810419}, 493, 661

\bibitem[\protect\citeauthoryear{Rodriguez-Merino, Chavez, Bertone  \&
  Buzzoni}{Rodriguez-Merino et~al.}{2005}]{rodriguez-merino2005}
Rodriguez-Merino L.~H.,  Chavez M.,  Bertone E.,   Buzzoni A.,  2005, \mn@doi
  [\apj] {10.1086/429858}, 626, 411

\bibitem[\protect\citeauthoryear{Rodriguez et~al.,}{Rodriguez
  et~al.}{2018}]{rodriguez2018}
Rodriguez J.~E.,  et~al., 2018, \mn@doi [\apj] {10.3847/1538-4357/aac08f}, 859,
  150

\bibitem[\protect\citeauthoryear{Rosotti, Dale, de Juan~Ovelar, Hubber,
  Kruijssen, Ercolano  \& Walch}{Rosotti et~al.}{2014}]{rosotti2014}
Rosotti G.~P.,  Dale J.~E.,  de Juan~Ovelar M.,  Hubber D.~A.,  Kruijssen J.
  M.~D.,  Ercolano B.,   Walch S.,  2014, \mn@doi [\mnras]
  {10.1093/mnras/stu679}, 441, 2094

\bibitem[\protect\citeauthoryear{Rosotti, Tazzari, Booth, Testi, Lodato  \&
  Clarke}{Rosotti et~al.}{2019}]{rosotti2019a}
Rosotti G.~P.,  Tazzari M.,  Booth R.~A.,  Testi L.,  Lodato G.,   Clarke C.,
  2019, \mn@doi [\mnras] {10.1093/mnras/stz1190}, 486, 4829

\bibitem[\protect\citeauthoryear{Sacco et~al.,}{Sacco et~al.}{2017}]{sacco2017}
Sacco G.~G.,  et~al., 2017, \mn@doi [\aap] {10.1051/0004-6361/201629698}, 601,
  A97

\bibitem[\protect\citeauthoryear{Scally \& Clarke}{Scally \&
  Clarke}{2001}]{scally2001}
Scally A.,  Clarke C.,  2001, \mn@doi [\mnras]
  {10.1046/j.1365-8711.2001.04274.x}, 325, 449

\bibitem[\protect\citeauthoryear{Scalo}{Scalo}{1990}]{scalo1990}
Scalo J.,  1990, \mn@doi [Physical Processes in Fragmentation and Star
  Formation] {10.1007/978-94-009-0605-1-12}, 162, 151

\bibitem[\protect\citeauthoryear{Sellek, Booth  \& Clarke}{Sellek
  et~al.}{2020}]{sellek2020}
Sellek A.~D.,  Booth R.~A.,   Clarke C.~J.,  2020, \mn@doi [\mnras]
  {10.1093/mnras/stz3528}, 492, 1279

\bibitem[\protect\citeauthoryear{Simon}{Simon}{1997}]{simon1997}
Simon M.,  1997, \mn@doi [\apjl] {10.1086/310678}, 482, L81

\bibitem[\protect\citeauthoryear{{Smallwood}, {Nealon}, {Chen}, {Martin}, {Bi},
  {Dong}  \& {Pinte}}{{Smallwood} et~al.}{2021}]{2021MNRAS.508..392S}
{Smallwood} J.~L.,  {Nealon} R.,  {Chen} C.,  {Martin} R.~G.,  {Bi} J.,  {Dong}
  R.,   {Pinte} C.,  2021, \mn@doi [\mnras] {10.1093/mnras/stab2624}, \href
  {https://ui.adsabs.harvard.edu/abs/2021MNRAS.508..392S} {508, 392}

\bibitem[\protect\citeauthoryear{Smith et~al.,}{Smith et~al.}{2020}]{smith2020}
Smith R.~J.,  et~al., 2020, \mn@doi [Monthly Notices of the Royal Astronomical
  Society] {10.1093/mnras/stz3328}, 492, 1594

\bibitem[\protect\citeauthoryear{St{\"o}rzer \& Hollenbach}{St{\"o}rzer \&
  Hollenbach}{1999}]{storzer1999}
St{\"o}rzer H.,  Hollenbach D.,  1999, \mn@doi [\apj] {10.1086/307055}, 515,
  669

\bibitem[\protect\citeauthoryear{Tobin et~al.,}{Tobin et~al.}{2020}]{tobin2020}
Tobin J.~J.,  et~al., 2020, \mn@doi [\apj] {10.3847/1538-4357/ab6f64}, 890, 130

\bibitem[\protect\citeauthoryear{{Toonen}, {Portegies Zwart}, {Hamers}  \&
  {Bandopadhyay}}{{Toonen} et~al.}{2020}]{2020A&A...640A..16T}
{Toonen} S.,  {Portegies Zwart} S.,  {Hamers} A.~S.,   {Bandopadhyay} D.,
  2020, \mn@doi [\aap] {10.1051/0004-6361/201936835}, \href
  {https://ui.adsabs.harvard.edu/abs/2020A&A...640A..16T} {640, A16}

\bibitem[\protect\citeauthoryear{Trapman, Rosotti, Bosman, Hogerheijde  \& van
  Dishoeck}{Trapman et~al.}{2020}]{trapman2020}
Trapman L.,  Rosotti G.,  Bosman A.~D.,  Hogerheijde M.~R.,   van Dishoeck
  E.~F.,  2020, \mn@doi [\aap] {10.1051/0004-6361/202037673}, 640, A5

\bibitem[\protect\citeauthoryear{Van Der~Walt, Colbert  \& Varoquaux}{Van
  Der~Walt et~al.}{2011}]{vanderwalt2011a}
Van Der~Walt S.,  Colbert S.~C.,   Varoquaux G.,  2011, \mn@doi [Computing in
  Science & Engineering] {10.1109/MCSE.2011.37}, 13, 22

\bibitem[\protect\citeauthoryear{Vicente \& Alves}{Vicente \&
  Alves}{2005}]{vicente2005}
Vicente S.~M.,  Alves J.,  2005, \mn@doi [\aap] {10.1051/0004-6361:20053540},
  441, 195

\bibitem[\protect\citeauthoryear{Vincke \& Pfalzner}{Vincke \&
  Pfalzner}{2016}]{vincke2016}
Vincke K.,  Pfalzner S.,  2016, \mn@doi [\apj] {10.3847/0004-637X/828/1/48},
  828, 48

\bibitem[\protect\citeauthoryear{Vincke \& Pfalzner}{Vincke \&
  Pfalzner}{2018}]{vincke2018}
Vincke K.,  Pfalzner S.,  2018, \mn@doi [\apj] {10.3847/1538-4357/aae7d1}, 868,
  1

\bibitem[\protect\citeauthoryear{Vincke, Breslau  \& Pfalzner}{Vincke
  et~al.}{2015}]{vincke2015}
Vincke K.,  Breslau A.,   Pfalzner S.,  2015, \mn@doi [\aap]
  {10.1051/0004-6361/201425552}, 577, A115

\bibitem[\protect\citeauthoryear{Virtanen et~al.,}{Virtanen
  et~al.}{2019}]{virtanen2019}
Virtanen P.,  et~al., 2019, arXiv:1907.10121 [physics]

\bibitem[\protect\citeauthoryear{Wall, McMillan, Mac~Low, Klessen  \&
  Portegies~Zwart}{Wall et~al.}{2019}]{wall2019}
Wall J.~E.,  McMillan S. L.~W.,  Mac~Low M.-M.,  Klessen R.~S.,
  Portegies~Zwart S.,  2019, \mn@doi [\apj] {10.3847/1538-4357/ab4db1}, 887, 62

\bibitem[\protect\citeauthoryear{Ward, Kruijssen  \& Rix}{Ward
  et~al.}{2020}]{ward2020}
Ward J.~L.,  Kruijssen J. M.~D.,   Rix H.-W.,  2020, \mn@doi [Monthly Notices
  of the Royal Astronomical Society] {10.1093/mnras/staa1056}, 495, 663

\bibitem[\protect\citeauthoryear{Williams \& Best}{Williams \&
  Best}{2014}]{williams2014}
Williams J.~P.,  Best W. M.~J.,  2014, \mn@doi [\apj]
  {10.1088/0004-637X/788/1/59}, 788, 59

\bibitem[\protect\citeauthoryear{Williams \& Cieza}{Williams \&
  Cieza}{2011}]{williams2011}
Williams J.~P.,  Cieza L.~A.,  2011, \mn@doi [\araa]
  {10.1146/annurev-astro-081710-102548}, 49, 67

\bibitem[\protect\citeauthoryear{Winter, Clarke, Rosotti  \& Booth}{Winter
  et~al.}{2018a}]{winter2018}
Winter A.~J.,  Clarke C.~J.,  Rosotti G.,   Booth R.~A.,  2018a, \mn@doi
  [\mnras] {10.1093/mnras/sty012}, 475, 2314

\bibitem[\protect\citeauthoryear{Winter, Clarke, Rosotti, Ih, Facchini  \&
  Haworth}{Winter et~al.}{2018b}]{winter2018a}
Winter A.~J.,  Clarke C.~J.,  Rosotti G.,  Ih J.,  Facchini S.,   Haworth
  T.~J.,  2018b, \mn@doi [\mnras] {10.1093/mnras/sty984}, 478, 2700

\bibitem[\protect\citeauthoryear{Winter, Booth  \& Clarke}{Winter
  et~al.}{2018c}]{winter2018b}
Winter A.~J.,  Booth R.~A.,   Clarke C.~J.,  2018c, \mn@doi [\mnras]
  {10.1093/mnras/sty1866}, 479, 5522

\bibitem[\protect\citeauthoryear{Winter, Clarke, Rosotti, Hacar  \&
  Alexander}{Winter et~al.}{2019}]{winter2019a}
Winter A.~J.,  Clarke C.~J.,  Rosotti G.~P.,  Hacar A.,   Alexander R.,  2019,
  \mn@doi [\mnras] {10.1093/mnras/stz2545}, 490, 5478

\bibitem[\protect\citeauthoryear{Winter, Kruijssen, Chevance, Keller  \&
  Longmore}{Winter et~al.}{2020a}]{winter2020a}
Winter A.~J.,  Kruijssen J. M.~D.,  Chevance M.,  Keller B.~W.,   Longmore
  S.~N.,  2020a, \mn@doi [\mnras] {10.1093/mnras/stz2747}, 491, 903

\bibitem[\protect\citeauthoryear{Winter, Kruijssen, Longmore  \&
  Chevance}{Winter et~al.}{2020b}]{winter2020b}
Winter A.~J.,  Kruijssen J. M.~D.,  Longmore S.~N.,   Chevance M.,  2020b,
  \mn@doi [\nat] {10.1038/s41586-020-2800-0}, 586, 528

\bibitem[\protect\citeauthoryear{Wright, Drake, Drew  \& Vink}{Wright
  et~al.}{2010}]{wright2010}
Wright N.~J.,  Drake J.~J.,  Drew J.~E.,   Vink J.~S.,  2010, \mn@doi [\apj]
  {10.1088/0004-637X/713/2/871}, 713, 871

\bibitem[\protect\citeauthoryear{Wright, Parker, Goodwin  \& Drake}{Wright
  et~al.}{2014}]{wright2014}
Wright N.~J.,  Parker R.~J.,  Goodwin S.~P.,   Drake J.~J.,  2014, \mn@doi
  [\mnras] {10.1093/mnras/stt2232}, 438, 639

\bibitem[\protect\citeauthoryear{de La~Fuente~Marcos \& de
  La~Fuente~Marcos}{de~La~Fuente~Marcos \&
  de~La~Fuente~Marcos}{2006}]{delafuentemarcos2006}
de La~Fuente~Marcos R.,  de La~Fuente~Marcos C.,  2006, \mn@doi [\aap]
  {10.1051/0004-6361:20054552}, 452, 163

\bibitem[\protect\citeauthoryear{van Terwisga, Hacar  \& van Dishoeck}{van
  Terwisga et~al.}{2019}]{vanterwisga2019}
van Terwisga S.~E.,  Hacar A.,   van Dishoeck E.~F.,  2019, \mn@doi [\aap]
  {10.1051/0004-6361/201935378}, 628, A85

\bibitem[\protect\citeauthoryear{van Terwisga et~al.,}{van Terwisga
  et~al.}{2020}]{vanterwisga2020}
van Terwisga S.~E.,  et~al., 2020, \mn@doi [\aap]
  {10.1051/0004-6361/201937403}, 640, A27

\makeatother
\end{thebibliography}

\bsp 
\label{lastpage}
\end{document}